\documentclass[twocolumn]{aastex63} 
\usepackage{amsmath}
\usepackage{multirow}
\usepackage{breakurl}
\usepackage{hyperref}
\usepackage{textcomp}
\usepackage{cleveref}[2012/02/15]
\usepackage{fancyhdr}
\usepackage{graphicx}
\usepackage{amssymb}
\usepackage{longtable}
\usepackage{float}
\usepackage{dcolumn}
\usepackage{enumitem}

\usepackage[toc,page]{appendix}
\usepackage[caption=false]{subfig}

\crefformat{footnote}{#2\footnotemark[#1]#3}

\newcommand{\kms}{km~s$^{-1}$}
\newcommand{\s}{$\sim$}
\newcommand{\n}{$-$}
\newcommand{\HI}{\ion{H}{1}}
\newcommand{\OI}{\ion{O}{1}}

\newcommand{\CII}{\ion{C}{2}}
\newcommand{\AlII}{\ion{Al}{2}}

\newcommand{\SiII}{\ion{Si}{2}}
\newcommand{\SiIII}{\ion{Si}{3}}

\newcommand{\FeII}{\ion{Fe}{2}}
\newcommand{\CIV}{\ion{C}{4}}

\newcommand{\SiIV}{\ion{Si}{4}}

\newcommand{\tm}{\tablenotemark} 
\newcommand{\tn}{\tablenotetext}

\shortauthors{Ashley, T. et al.}

\begin{document}

\title{The Metallicities of Five Small High-Velocity Clouds\footnote{Based on observations made with the NASA/ESA Hubble Space Telescope, obtained from the Data Archive at the Space Telescope Science Institute, which is operated by the Association of Universities for Research in Astronomy, Inc., under NASA contract NAS5-26555. These observations are associated with program 15887.}}

\author[0000-0002-6541-869X]{Trisha Ashley}
\affiliation{Space Telescope Science Institute, 3700 San Martin Drive, Baltimore, MD 21218}

\author[0000-0003-0724-4115]{Andrew J. Fox}
\affiliation{AURA for ESA, Space Telescope Science Institute, 3700 San Martin Drive, Baltimore, MD 21218}
\affiliation{Department of Physics \& Astronomy, Johns Hopkins University, 3400 N. Charles Street, Baltimore, MD 21218, USA}
\email{afox@stsci.edu}

\author[0000-0002-6050-2008]{Felix J. Lockman}
\affiliation{Green Bank Observatory, P.O. Box 2, Rt. 28/92, Green Bank, WV 24944, USA}

\author[0000-0002-0507-7096]{Bart P. Wakker}
\affiliation{Department of Astronomy, University of Wisconsin-Madison, 475 North Charter Street, Madison, WI 53706, USA}

\author[0000-0002-1188-1435]{Philipp Richter}
\affiliation{Institut für Physik und Astronomie, Universität Potsdam, Haus 28, Karl-Liebknecht-Str. 24/25, D-14476 Golm (Potsdam), Germany}

\author[0000-0003-3681-0016]{David M. French}
\affiliation{Space Telescope Science Institute, 3700 San Martin Drive, Baltimore, MD 21218}

\author[0000-0002-3005-9738]{Vanessa A. Moss}
\affiliation{CSIRO Space \& Astronomy, Marsfield, New South Wales, Australia}
 
\author[0000-0003-2730-957X]{Naomi M. McClure-Griffiths}
\affiliation{Research School of Astronomy and Astrophysics, The Australian National University, Canberra, Australian Capital Territory 2611, Australia}

\begin{abstract}
High-velocity clouds (HVCs) are  multi-phase gas structures
whose velocities ($|v_{\rm LSR}|\ge100$ \kms) are too high to be 
explained by Galactic disk rotation. 
While large HVCs are well characterized, compact and small HVCs 
(with \HI\ angular sizes of a few degrees) are poorly understood.
Possible origins for such small clouds include Milky Way halo gas or fragments of the Magellanic System, but neither their origin nor their connection to the Milky Way halo has been confirmed.  We use new Hubble Space Telescope/Cosmic Origins Spectrograph UV spectra and Green Bank Telescope \HI\ spectra to measure the metallicities of five small HVCs in the southern Galactic sky projected near the Magellanic System. We build a set of distance-dependent \textit{Cloudy} photoionization models for each cloud and calculate their ionization-corrected metallicities. 
All five small HVCs have oxygen metallicities $\le$0.17 $Z_\sun$, indicating they do not originate in the disk of the Milky Way. Two of the five have metallicities of 0.16--0.17 $Z_\sun$, similar to the Magellanic Stream, suggesting these clouds are fragments of the Magellanic System. The remaining three clouds have much lower metallicities of 0.02--0.04 $Z_\sun$. 
While the origin of these low-metallicity clouds is unclear, they could be gaseous mini-halos or gas stripped from 
dwarf galaxies by ram pressure or tidal interactions. These results suggest that small HVCs do not all reside in the inner 
Milky Way halo or the Magellanic System, but instead can trace more distant structures. 
\end{abstract}


\section{Introduction}

High-velocity clouds (HVCs) are multi-phase gas clouds observed at velocities too large to be accounted for by disk rotation, typically $|v_{\rm LSR}|\ge100$ \kms\ \citep{Wakker_1997}. HVCs play an integral role in regulating the gaseous ecosystem of the Milky Way (MW) and its star formation rate \citep{Putman_2012, Richter_2017, Fox_2019}. Their inflow to the MW disk provides new fuel for future star formation, while their outflow reduces the available gas supply. \citet{Fox_2019} and \citet{French_2021} use HVC UV data to show that the MW is currently in an inflow-dominated phase, with an HVC inflow rate of $\ga$0.53$\pm$0.23 M$_{\sun}$ yr$^{-1}$. While this inflow rate is not sufficient to sustain the current star formation rate of $1.7\pm0.7$ M$_{\sun}$ yr$^{-1}$ \citep{Chomiuk_2011, Licquia_2015}, HVCs still play an important role in replenishing the MW gaseous disk with fresh gas for future generations of star formation.

A well-known source of HVCs in the southern Galactic hemisphere is the Magellanic System. The System contains the Large and Small Magellanic Clouds (LMC and SMC), the Magellanic Stream of stripped gas trailing the Clouds, the Magellanic Bridge of gas connecting the Clouds, and a gaseous Leading Arm in front of the Clouds \citep[see][and references therein]{Putman_1998, Putman_2003, Bruns_2005, Nidever_2008, Nidever_2010, DOghia_2016}. HVCs belonging to the Magellanic System have been identified by their projected positions, position-velocity diagrams, and low metallicities of typically $0.10-0.17$ $Z_\sun$ \citep{Lu_1998, Gibson_2001, Fox_2010, Fox_2013, Fox_2018, Richter_2013, Richter_2018}; albeit the metallicity of the LMC filament of the Stream is $\approx0.50$ $Z_\sun$ \citep{Gibson_2000, Richter_2013}. While the Magellanic System dominates the HVC population in the southern Galactic hemisphere, there are a significant number of HVCs that are not clearly associated with the Magellanic System and whose origin is unknown \citep{Richter_2001, Putman_2012,  Westmeier_2017}. 

A poorly-understood subset of HVCs are the Compact HVCs (CHVCs), defined as isolated clouds with \HI\ diameters on the sky of $<2\degr$ \citep{Braun_1999, deHeij_2002a, deHeij_2002b, deHeij_2002c, Putman_2002, Saul_2012}. 
Many small HVCs also exist that are not isolated or are slightly larger than $2\degr$; these are not formally considered CHVCs, but their origin is equally unknown.
CHVCs have been extensively studied in \HI, while infrared searches for stellar components in CHVCs have not been successful, making it difficult to put strong constraints their distances \citep{Braun_1999, Hopp_2003, Hopp_2006}. Distance estimates of CHVCs from \HI\ observations based on kinematics, thermal pressure, comparisons to known dwarf galaxies, the spatial distribution of CHVCs, and the virial theorem yield distances of \s100-850 kpc, placing the CHVCs firmly within the Local Group \citep{Braun_2000,Bruns_2001, Burton_2001, Sternberg_2002, deHeij_2002c, Maloney_2003, Westmeier_2005, Pisano_2004, Pisano_2007}.

Compact and small HVCs have many possible origins, including:
(1) gas ejected from the Galactic disk in an energetic outflow, 
(2) extragalactic gas infalling through the Galactic halo, 
(3) fragments of the Magellanic System, 
(4) ultra-faint Local Group dwarf galaxies whose stars are yet to be detected or mini-halos, small dark-matter halos with gas and no star formation,  
(5) gas stripped from Local Group dwarf galaxies via tidal interactions or ram pressure, and
(6) intergalactic gas \citep{Oort_1966, Giovanelli_1978, Bregman_1980, Braun_1999, Blitz_1999, Bruns_2000, Bruns_2001, Sternberg_2002, deHeij_2002c, Pisano_2004, Pisano_2007, Putman_2006, Putman_2012}. For example, five CHVCs have been identified by \citet{Nidever_2010} as Magellanic Stream (MS) components based on their continuity in position-velocity diagrams with the MS. However, CHVCs are found across the sky and therefore cannot always be attributed to Magellanic gas \citep{Putman_2012, Moss_2017}.

UV absorption-line studies can be used to measure the ionized gas content, metallicities, and dust depletion levels of 
compact and small HVCs and therefore constrain their origin. Until now, UV absorption studies have been conducted for 
only two CHVCs: CHVC 224.0$-$83.4$-$197, by \citet{Sembach_2002}, \citet{Richter_2009}, and \citet{Kumari_2015} and CHVC
125+41$-$207, by \citet{Bowen_1993} and \citet{Braun_2000}. CHVC 224.0-83.4-197 is located in the southern Galactic 
hemisphere \s$10$\degr\ from the \HI\ MS, and is probed by the sightline to the QSO Ton\,S210. \citet{Kumari_2015} 
determined a low metallicity of [O/H] = $-1.12\pm0.22$ ($0.076$ $Z_\sun$) for this CHVC and concluded it is likely 
a fragment of the MS. CHVC 125+41$-$207 is located in the northern Galactic hemisphere and has a measured metallicity of 
0.04--0.07 $Z_\sun$. \citet{Braun_2000} suggest that this measurement along with \HI\ kinematics of CHVC 125+41$-$207 
imply a Local Group dark-matter-dominated self-gravitating object, possibly a low surface brightness dwarf galaxy. 
These are valuable measurements, but more a larger sample of metallicities is needed to draw general conclusions.

In this paper, we present the first systematic study of the metallicities of small HVCs using a sample of five southern Galactic hemisphere clouds identified from the Galactic All-Sky Survey (GASS) \citep{McClure-Griffiths_2009, Moss_2013, Moss_2017}. The UV and \HI\ observations and data reduction are presented in Section~\ref{section:observations}. In Section~\ref{section:results} we present the calculations for each HVC's ionization-corrected gas-phase abundances and dust depletion levels. We discuss potential explanations for the HVC metal abundances 
in Section~\ref{section:discussion}. We then summarize our results in Section~\ref{section:summary}.

\section{Observations and Data Reduction}\label{section:observations}

\begin{deluxetable*}{lccccccc}[!ht]
\tablecaption{Basic Sight Line \& HVC Information\label{table:basic_info}}
\tablehead{\colhead{Sight line} &  \colhead{$l$} & \colhead{$b$}  & \colhead{GASS HVC\tm{a}} & \colhead{HIPASS HVC\tm{b}} & \colhead{$v_{\mathrm{LSR}}$\tm{c}} & \colhead{$\Delta$RA\tm{d}} & \colhead{$\Delta$Dec\tm{d}}\\[-6pt]
\colhead{} & \colhead{(\degr)} & \colhead{(\degr)} & \colhead{} &
& \colhead{(\kms)}  & \colhead{(\degr)} & \colhead{(\degr)} }
\startdata
CTS\,47 &  $236.7$ & $-40.9$  &  G237.2–41.1+146 & 236.5-40.6+160 & $146.7\pm2.6$ & 1.4 & 2.5 \\ 
HE\,0027  & $352.3$ & $-83.8$  &  G348.3–83.8–192 &   347.1-83.8-192 & $-192.8\pm3.1$ & 1.0 & 1.6 \\
IRAS\,0459  & $223.5$ & $-33.7$ &   G224.0–34.3+135 &   224.1-34.4+136  & $135.6\pm1.0$ & 1.5 & 2.1 \\ 
Mrk\,969 & $133.9$ & $-75.5$ &  G133.5–75.6–294 &   131.9-75.8-302  & $-294.1\pm1.8$ & 1.6 & 1.1\\ 
UVQS\,J0110  & $145.6$ & $-77.8$ & G146.2–77.6–279 &   146.4-77.8-277 & $-279.7\pm1.5$  
 & 0.9 & 1.6 \\ 
\enddata
\tn{a}{GASS HVC designations are defined by longitude-latitude-velocity \citep{Moss_2013}.}
\tn{b}{HIPASS HVC designations \citep{Putman_2002}.}
\tn{c}{Local Standard of Rest (LSR) velocity of the HVC, measured by \citet{Moss_2013}.}
\tn{d}{HVC angular size in Right Ascension and Declination within the \HI\ contours shown in Figure~\ref{figure:chvc_hi}.}
\end{deluxetable*}

\begin{figure*}[!ht]
    \epsscale{1.1}
      \plotone{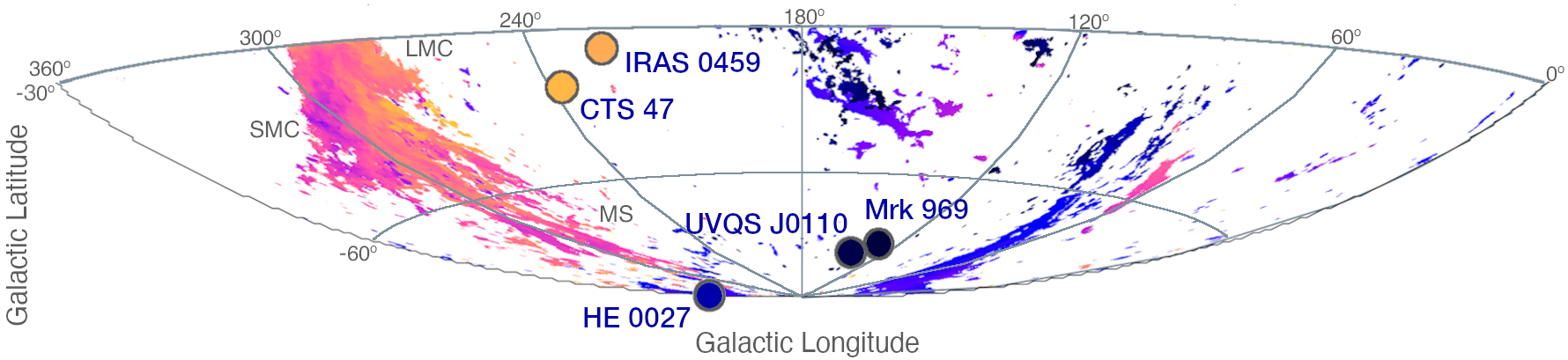}
      \plotone{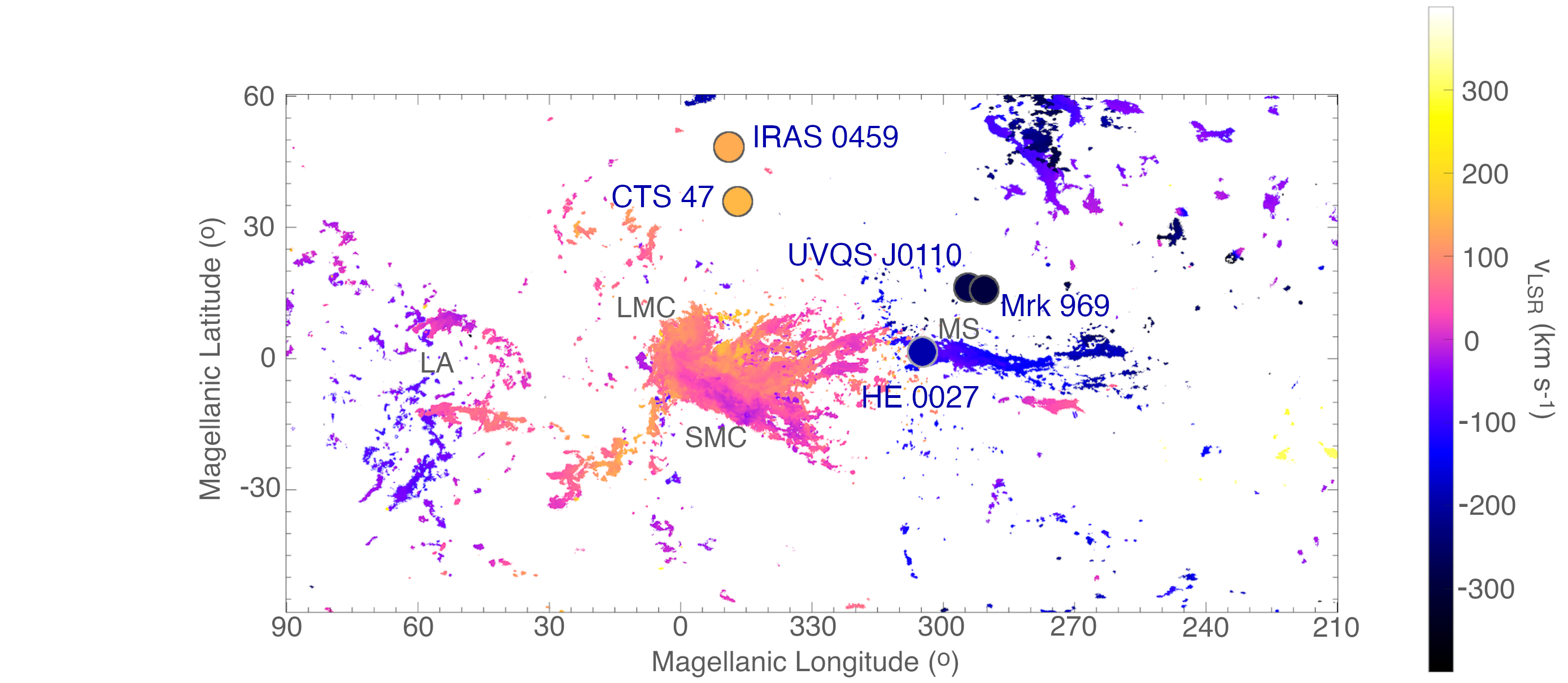}
\caption{\HI\ 21\,cm emission maps of HVCs in Galactic and Magellanic coordinates from HI4PI data \citet{Bekhti_2016, Westmeier_2017}, showing the relative location of the five HVCs (and their background AGN) in this work and their position with respect to the Magellanic System. The AGN are shown as circles. The colors of the AGN circles and the \HI\ emission represent the LSR velocities of the high-velocity absorption/emission.} 
\label{figure:westmeier}
\end{figure*}

\begin{figure*}[!ht]
    \centering
    \includegraphics[width=0.3\textwidth]{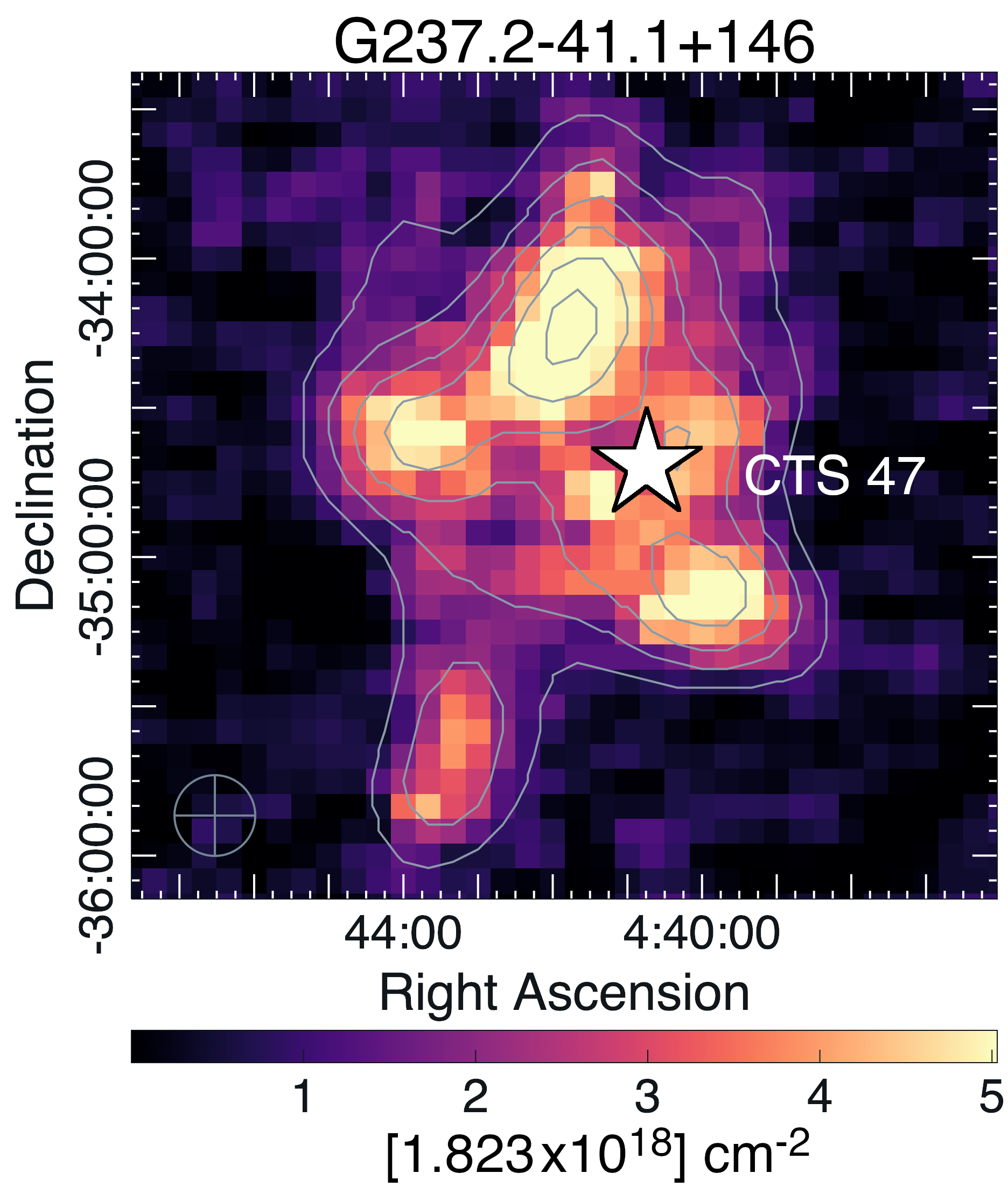}  
    {\includegraphics[width=0.27\textwidth]{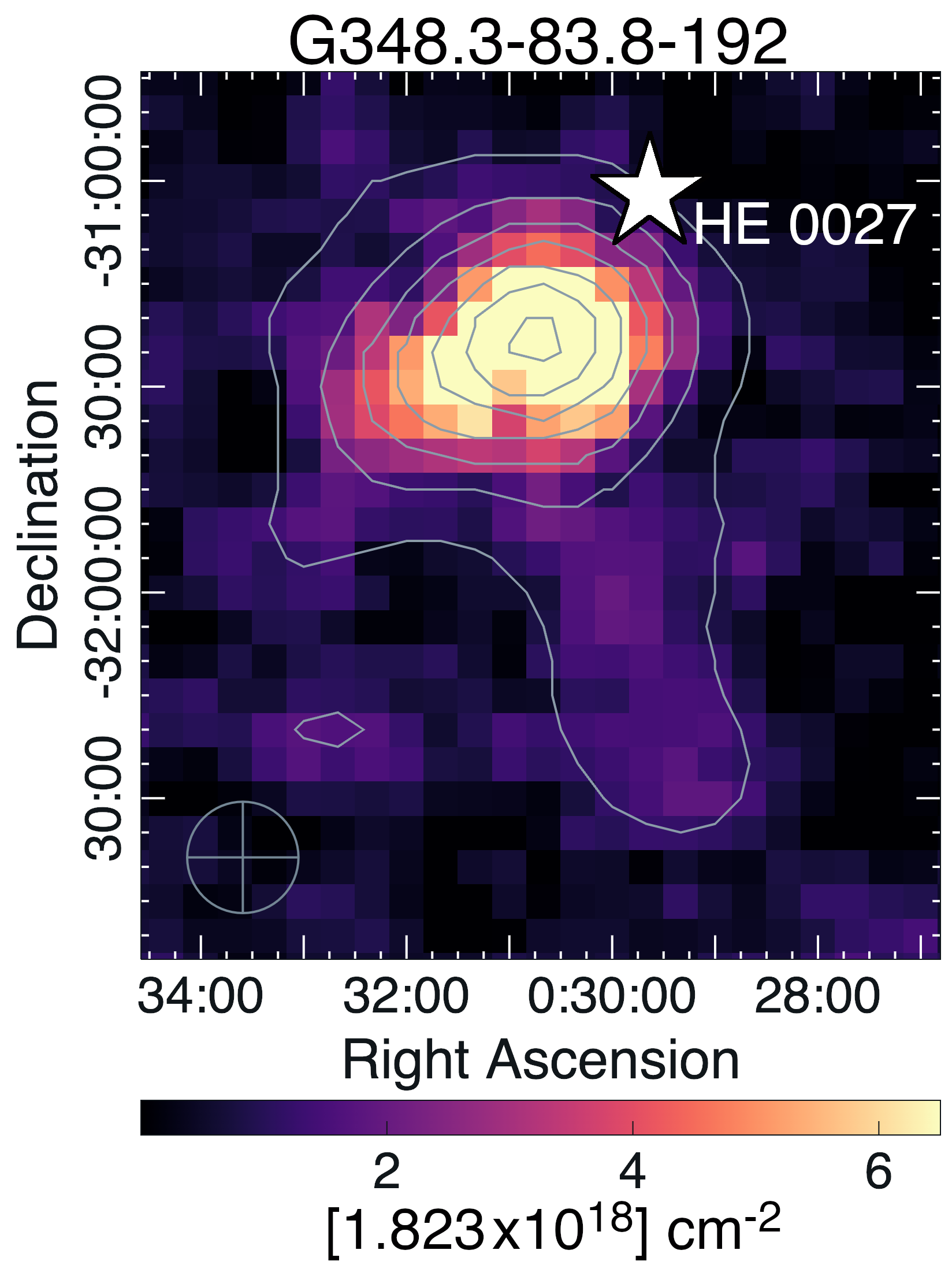}} 
    \includegraphics[width=0.25\textwidth]{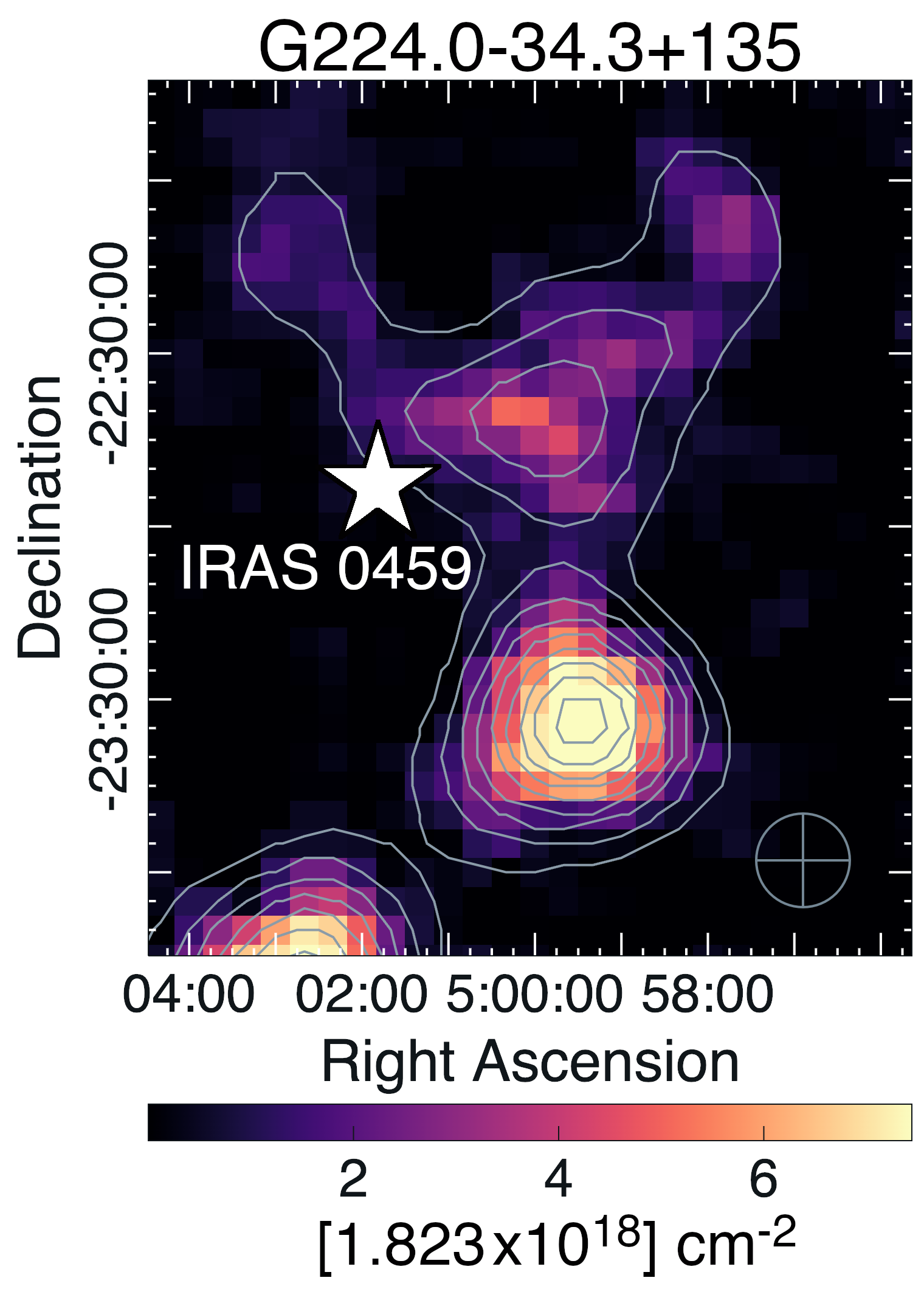} \\[12pt]
    \begin{tabular}{cc}
   \raisebox{0.15\height}
    {\includegraphics[width=0.35\textwidth]{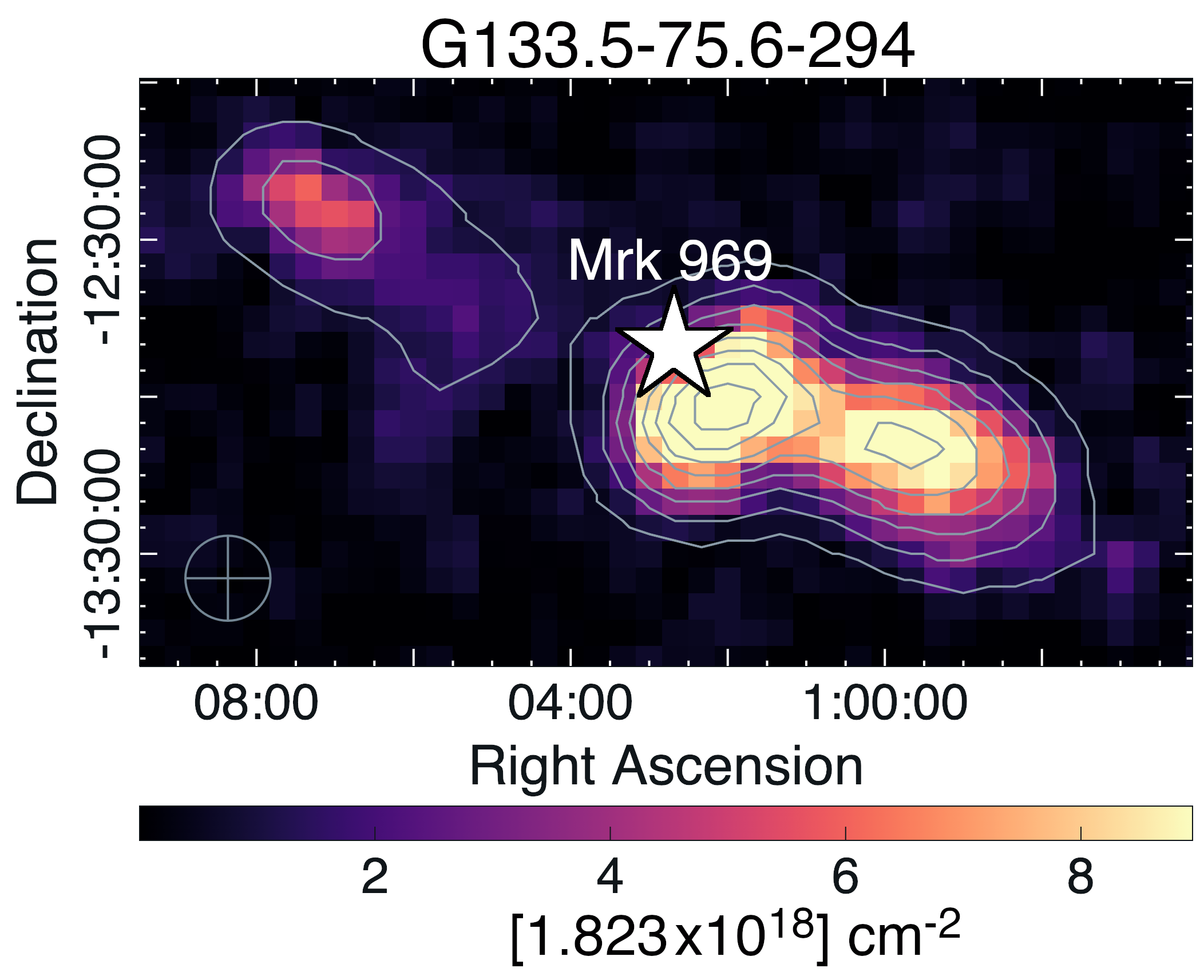}} &
    \includegraphics[width=0.23\textwidth]{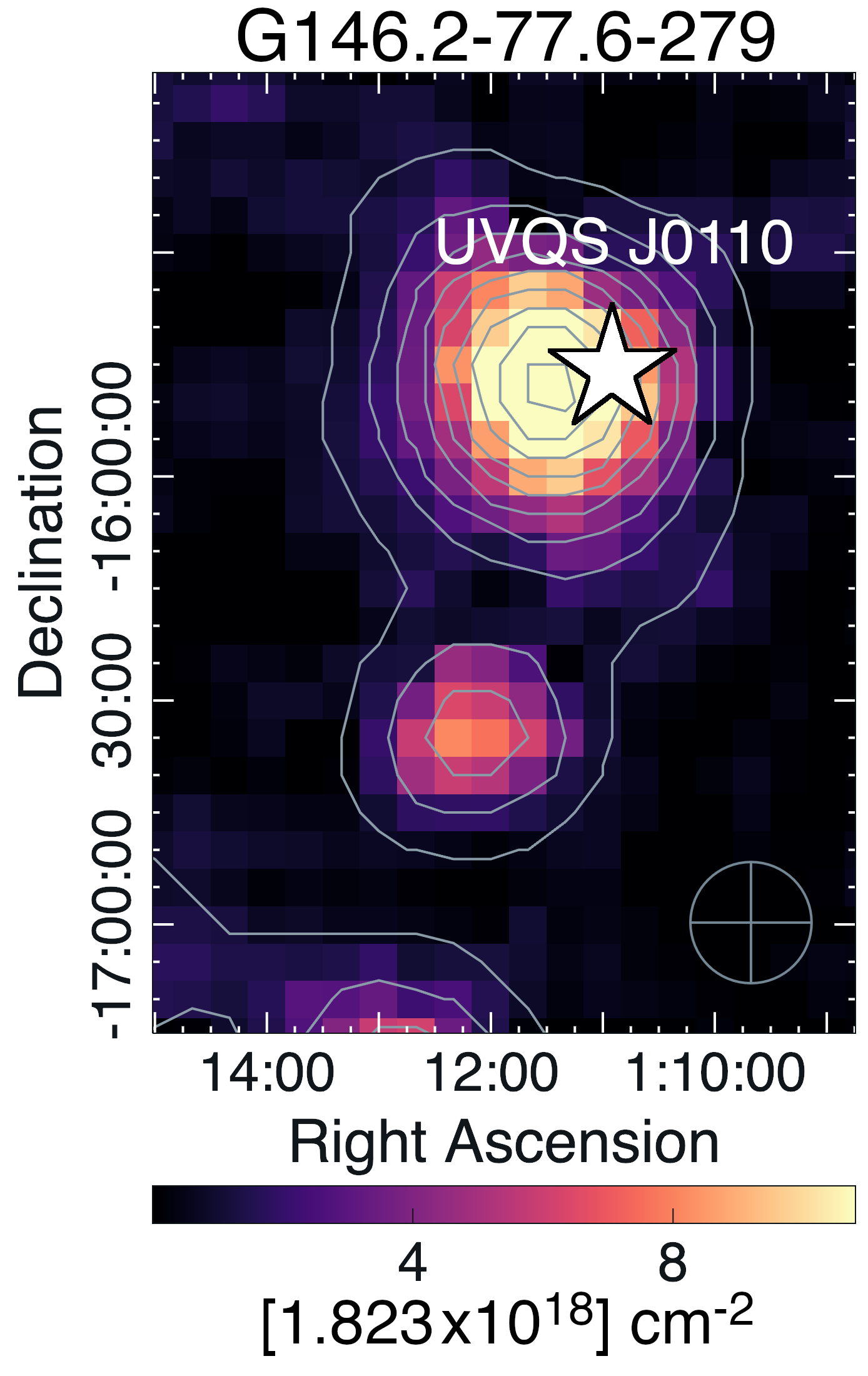}\\
    \end{tabular}
\caption{\HI\ 21cm intensity maps of the small HVC sample using HI4PI data \citep{Bekhti_2016}. The stars represent the projected location of the background AGN. 
The contours shown are at the following $N$(\HI) levels.
G237.2-41.1+146: (2.3, 3.8, 5.3, 6.8, 8.3, 9.8)$\times$10$^{18}$ cm$^{-2}$;
G348.3-83.8-192: (1.9, 3.8, 5.7, 7.6, 9.5, 11.4, 13.3)$\times$10$^{18}$ cm$^{-2}$; 
G224.0-34.3+135: (1.4, 2.8, 4.2, 5.6, 7.0, 8.4, 9.8, 11.2, 12.6)$\times$10$^{18}$ cm$^{-2}$; 
G133.5-75.6-294: (2.3, 4.6, 6.9, 9.2, 11.5, 13.8, 16.1, 18.4)$\times$10$^{18}$ cm$^{-2}$; 
G146.2-77.6-279: (2.2, 4.4, 6.6, 8.8, 11.0, 13.2, 15.4, 17.6, 19.8)$\times$10$^{18}$ cm$^{-2}$.} 
\label{figure:chvc_hi}
\end{figure*}

To form our sample, we cross-correlated the GASS HVC catalog \citep{Moss_2013} with a compiled list of UV-bright AGN, 
and found five HVCs with \HI\ diameters on the sky of $\lesssim$2--3$\degr$ with a UV-bright AGN projected behind them.
These five HVCs are located in the southern Galactic hemisphere as shown in Figure~\ref{figure:westmeier}. 
Close-up \HI\ maps of the clouds are shown in Figure~\ref{figure:chvc_hi}.
These clouds are not all strictly considered as CHVCs based on the original definition of \citet{Braun_1999},
because two have sizes just above the 2$\degr$ threshold, and two others are not isolated, 
which is a secondary factor in identifying CHVCs; 
instead, we refer to the five clouds as ``small HVCs" throughout this analysis.
The HVCs all have identifications in the original HIPASS catalog of HVCs \citep{Putman_2002} as well as GASS identifications.
Each of the five background AGN lies directly behind their associated HVC \HI\ emission and less than 1\degr\ from the peak GASS \HI\ flux (see Figure~\ref{figure:chvc_hi}). 
See Table~\ref{table:basic_info} for basic information on each sight line and HVC, including their GASS and HIPASS identifications.
For brevity, we adopt abbreviated names for the following background AGN: HE\,0027 = HE\,0027-3118, IRAS\,0459 = IRAS\,04596-2257, and UVQS\,J0110 = UVQS\,J011054.99-154540.2. 

\subsection{UV Observations}

Each target AGN was observed with the Hubble Space Telescope Cosmic Origins Spectrograph (HST/COS) under Cycle 27 Program 15887 (PI: A. Fox). The spectra were taken using both the G130M/1222 and G160M/1533 grating settings using four FP-POS positions per setting. The data were reduced with the \texttt{calcos} pipeline twice, once using all photons collected and a second time using ``night-only'' photons to reduce geocoronal emission, which can strongly contaminate \OI\ $\lambda$1302 absorption, a key line for our metallicity analysis. A customized set of velocity alignment and co-addition steps was then applied to the data 
following the procedure outlined in \citet{Wakker_2015}. 

\begin{figure*}[!ht]
    \centering
    \epsscale{1}
      \plotone{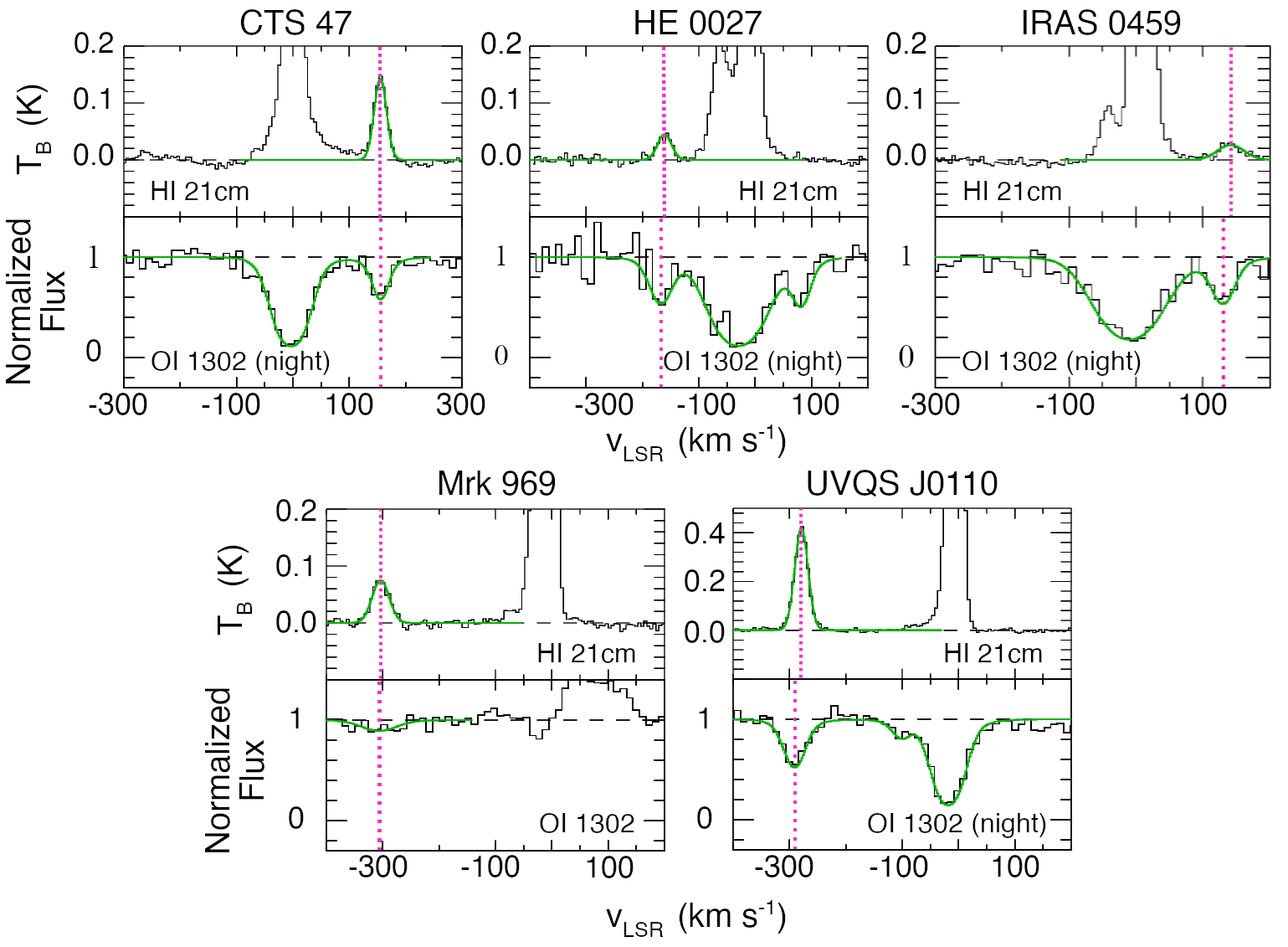}
\caption{HST/COS \OI\ absorption-line spectra and GBT \HI\ emission-line spectra used for the HVC metallicity measurements. The green lines show Voigt fits to the UV absorption and Gaussian fits to the HVC \HI\ emission. The pink vertical dotted lines represent the velocity centroids of the \HI\ emission and the \OI\ absorption.}\label{figure:cts-uv-fits}
\end{figure*}

We normalized the spectra around each line of interest using polynomial fits to regions of unabsorbed continuum. We then used the VPFIT software package \citep{Carswell_2014} to identify and fit Voigt profiles to the HVC absorption features using a chi-squared minimization technique. To be considered real, the absorption component was required to appear in two or more metal ions. The data were then binned by five pixels and visually inspected. Summary plots of the \OI\ absorption profiles and \HI\ emission profiles used in the metallicity calculations are shown in Figure~\ref{figure:cts-uv-fits}, with their associated fit parameters listed in Table~\ref{table:fits}. Fit parameters of all metal lines detected in the HVCs are given in Table~\ref{table:VPFIT}. Plots of the metal-line absorption and their associated fits are shown in Appendix~\ref{appendix:UV_absorption}.

\begin{deluxetable*}{lccccccccc}[!ht]
\tabletypesize{\small}
\tablecaption{Fits to Radio and UV Spectral Lines used for Metallicity Calculations\label{table:fits}}
\tablehead{\colhead{Sight line}  &  \colhead{Corotation} & \multicolumn{3}{c}{\underline{21cm Emission Gaussian Fit Parameters}} &  \multicolumn{4}{c}{\underline{~~~~~~~~UV Absorption Voigt Fit Parameters~~~~~~~~}} \\[-6pt]
\colhead{} &  \colhead{Velocities\tm{\scriptsize a}} &  \colhead{$v_{0\: \mathrm{HI}}$}  & \colhead{$b$-value} &   \colhead{log $N$} & \colhead{X$^{i}$} & \colhead{$v_{0\: \mathrm{UV}}$\tm{\scriptsize b}} &  \colhead{$b$-value} &  \colhead{log $N$} \\[-6pt]
\colhead{} &  \colhead{(\kms)} & \colhead{(K)} & \colhead{(\kms)}  &  \colhead{($N$ in cm$^{-2}$)} &  \colhead{} & \colhead{(\kms)} & \colhead{(\kms)} & \colhead{($N$ in cm$^{-2}$)}}
\startdata
CTS\,47    & $0:21$ & $155.2\pm0.2$ & $15.6\pm0.3$ & $18.85\pm0.01$ & \OI & $155.9\pm7.6$ & $11.6\pm3.1$ & $14.20\pm0.10$  \\ 
HE\,0027    & $0:0$ & $-160.8\pm1.0$ & $17.5\pm1.4$ & $18.40\pm0.05$ & \OI & $-166.9\pm8.9$ & $19.6\pm0.0$\tm{\scriptsize c} 
&  $14.33\pm0.09$  \\ 
IRAS\,0459   & $0:29$ & $141.7\pm1.8$ & $28.4\pm2.5$ & $18.38\pm0.05$ & \OI & $130.8\pm8.1$ & $15.8\pm5.5$ & $14.30\pm0.07$    \\	 
Mrk\,969  &  $-2:0$  & $-303.7\pm0.5$ & $21.4\pm0.8$ & $18.72\pm0.02$ & \OI & $-305.8\pm11.2$ & $38.0\pm0.0$\tm{\scriptsize c} & $13.74\pm0.08$\tm{\scriptsize d} \\ 
UVQS\,J0110  & $-1:0$ & $-278.9\pm0.1$ & $16.8\pm0.1$ & $19.36\pm0.00$ & \OI & $-290.4\pm7.8$ & $17.9\pm3.8$ & $14.35\pm0.05$ \\ 	
\enddata
\tn{a}{The allowed velocity range of gas corotating with the MW disk, calculated using methods described in \citet{Wakker_1991}.}
\tn{b}{Error includes the 7.5 \kms\ COS zero-point offset \citep{Plesha_2019}.}
\tn{c}{These fits used a fixed $b$-value to match that of the \SiII\ for HE0027 and \CII\ Mrk 969 (see Table~\ref{table:VPFIT}).}
\tn{d}{Geocoronal \OI\ emission does not contaminate the velocity range of the HVC towards Mrk 969. Therefore, this fit was made using the combined (day+night) spectrum. All other \OI\ measurements use night-only spectra.}
\end{deluxetable*}

The \OI\ $\lambda$1302 HVC absorption towards HE\,0027 (HVC G348.3–83.8–192) is partially blended with MW ISM absorption and has a reduced S/N ratio due to the night-only reduction. These effects made it difficult for VPFIT to obtain a fit for the HVC \OI\ absorption. We tested the fit by requiring the \OI\ linewidth ($b$-value) to match that of either the \SiII\ at 19.6 \kms\ or \CII\ at 32.6 \kms. The \SiII\ $b$-value resulted in a lower chi-squared value and a better visual match to the \OI\ absorption than the \CII\ $b$-value, so we adopt $b$=19.6 \kms\ for the \OI\ absorption measurement for the HVC toward HE\,0027.

Due to the reduced S/N ratio in the night-only spectrum of Mrk 969, VPFIT was unable to produce a reasonable fit to the \OI\ $\lambda$1302 absorption in HVC G133.5–75.6–294. Fortunately, the geocoronal \OI\ emission in the combined (day+night) spectrum of Mrk 969 is centered at positive velocities while HVC G133.5–75.6–294 has a central velocity of $-$294.1 \kms\ (see Appendix~\ref{appendix:UV_absorption}), far enough away in velocity to be uncontaminated by the airglow. As such, in this sightline we were able to use the combined spectrum to determine a fit to the HVC \OI\ absorption, which is significant at the 3.3$\sigma$ level.

For the HVC toward UVQS\,J0110 (HVC G146.2–77.6–279), the \FeII\ Voigt fit is based on both $\lambda$1144 and $\lambda$1608, but has a fixed $b$-value based on a Voigt fit to only \FeII\ $\lambda$1608, because VPFIT could not find a good fit to both lines simultaneously with $b$ as a free parameter. \FeII\ $\lambda$1144 is the stronger of the two lines, but it appears weaker in the absorption profile. Therefore, we conservatively set UVQS\,J0110's \FeII\ column density to an upper limit.

\begin{deluxetable*}{lcccc}[!ht]
\tablecaption{Voigt Profile Fits to HVC Absorption}\label{table:VPFIT}
\tabletypesize{\footnotesize}
\tablehead{\colhead{Sight Line}  & \colhead{Transition ($\lambda$)}  & \colhead{$v_0$ (\kms)} & \colhead{$b$ (\kms)} & \colhead{log\,$N$ ($N$ in cm$^{-2}$)}}
\startdata
CTS\,47 &  \ion{O}{1} $1302$ & $156\pm8$ & $12\pm3$ & $14.22\pm0.06$\\
  & \ion{N}{1} 1199, 1200.2, 1200.7  & ... & ... & $\le13.82$\tm{a}\\
 &  \ion{C}{2} $1334$ & $158\pm8$ & $22\pm3$ & $14.04\pm0.03$\\
 &  \ion{Al}{2} $1670$ & $171\pm10$ & $17\pm11$ & $12.23\pm0.13$\\
 &  \ion{Si}{2} $1190$,$1193$,$1260$ & $161\pm8$ & $14\pm2$ & $13.19\pm0.05$\\
 &  \ion{Si}{3} $1206$ & $174\pm14$ & $44\pm17$ & $12.70\pm0.12$\\
   & \ion{S}{2} 1250, 1253, 1259  & ... & ... & $\le14.50$\tm{a}\\
   & \ion{Fe}{2} $1608$,$1144$ & ... & ... & $\le13.96$\tm{a}\\
 &  \ion{C}{4} $1550$,$1548$ & ... & ... & $\le13.61$\tm{a}\\ 
 &  \ion{Si}{4} $1393$,$1402$ & ... & ... & $\le12.96$\tm{a}\\ \hline
HE\,0027 &  \ion{O}{1} $1302$ & $-167\pm9$ & $20\pm0$\tm{b} & $14.33\pm0.09$ \\
  & \ion{N}{1} 1199, 1200.2, 1200.7  & ... & ... & $\le13.23$\tm{a}\\
 & \ion{C}{2} $1334$ & $-153\pm8$ & $33\pm2$  & $14.53\pm0.03$\\
 & \ion{Al}{2} $1670$ & $-156\pm8$ &  $13\pm6$ & $12.90\pm0.23$ \\
 & \ion{Si}{2} $1190$,$1193$,$1260$ & $-157\pm8$ & $20\pm1$ & $13.81\pm0.03$\\
 & \ion{Si}{3} $1206$ & $-157\pm10$ & $19\pm6$ & $\ge13.51$\\
   & \ion{S}{2} 1250, 1253, 1259  & ... & ... & $\le14.10$\tm{a}\\
 & \ion{Fe}{2} $1608$,$1144$ & $-162\pm8$ & $14\pm6$ & $13.81\pm0.07$\\
 & \ion{C}{4} $1550$,$1548$ & $-140\pm8$ & $33\pm5$ & $13.93\pm0.04$\\
 & \ion{Si}{4} $1393$,$1402$ & $-157\pm8$ & $21\pm1$ & $13.46\pm0.01$\\ \hline
IRAS\,0459 & \ion{O}{1} $1302$ & $131\pm8$ & $16\pm6$ & $14.30\pm0.07$\\
 & \ion{N}{1} 1199, 1200.2, 1200.7  & ... & ... & $\le13.88$\tm{a}\\
 & \ion{C}{2} $1334$ & $153\pm8$ & $43\pm3$ & $14.37\pm0.02$\\
 & \ion{Al}{2} $1670$ & $144\pm8$  & $18\pm5$ & $12.83\pm0.09$\\
 & \ion{Si}{2} $1190$,$1193$,$1260$ & $148\pm8$ & $28\pm1$ & $13.40\pm0.01$\\
 & \ion{Si}{3} $1206$ & $146\pm9$ & $46\pm7$ & $\ge13.45$\\
   & \ion{S}{2} 1250, 1253, 1259  & ... & ... & $\le14.09$\tm{a}\\
  & \ion{Fe}{2} $1608$,$1144$ & ... & ... & $\le13.76$\tm{a}\\
 & \ion{C}{4} $1548$,$1550$ & ... & ... & $\le13.77$\tm{a}\\ 
  &  \ion{Si}{4} $1393$,$1402$ & ... & ... & $\le12.90$\tm{a}\\ \hline
Mrk\,969 & \ion{O}{1} $1302$ & $-306\pm11$ & $38\pm0\tm{b}$ & $13.74\pm0.08$\\
  & \ion{N}{1} 1199, 1200.2, 1200.7  & ... & ... & $\le13.71$\tm{a}\\
& \ion{C}{2} $1334$ & $-297\pm8$ & $38\pm4$ & $13.91\pm0.03$\\
 & \ion{Al}{2} $1670$ & $-296\pm9$ & $21\pm8$ & $12.53\pm0.09$\\
 & \ion{Si}{2} $1190$,$1193$,$1260$  & $-296\pm8$ & $35\pm3$ & $13.09\pm0.03$\\
 & \ion{Si}{3} $1206$ & $-298\pm8$ & $39\pm4$ & $13.03\pm0.03$\\
 & \ion{S}{2} 1250, 1253, 1259  & ... & ... & $\le14.30$\tm{a}\\
  & \ion{Fe}{2} $1608$,$1144$ & ... & ... & $\le13.98$\tm{a}\\
 & \ion{C}{4} $1548$,$1550$ & $-291\pm11$ & $43\pm11$ & $13.53\pm0.08$\\ 
   &  \ion{Si}{4} $1393$,$1402$ & ... & ... & $\le13.18$\tm{a}\\ \hline
 UVQS\,J0110 & \ion{O}{1} $1302$ & $-290\pm8$ & $18\pm4$ & $14.35\pm0.05$\\
  & \ion{N}{1} 1199, 1200.2, 1200.7  & ... & ... & $\le13.22$\tm{a}\\
 & \ion{C}{2} $1334$ & $-272\pm8$ & $24\pm3$ & $13.93\pm0.03$\\
 & \ion{Al}{2} $1670$ & ... & ... & $\le12.52$\tm{a}\\
 & \ion{Si}{2}\ $1190$,$1193$,$1260$  & $-274\pm8$ & $19\pm3$ & $13.18\pm0.04$\\
 & \ion{Si}{3} $1206$ & $-284\pm9$ & $29\pm6$ & $12.90\pm0.06$\\
   & \ion{S}{2} 1250, 1253, 1259  & ... & ... & $\le14.10$\tm{a}\\
 & \ion{Fe}{2} $1608$,$1144$ & ... & ... & $\le13.94$\tm{a}\\
  &  \ion{C}{4} $1550$,$1548$ & ... & ... & $\le13.04$\tm{a}\\ 
 &  \ion{Si}{4} $1393$,$1402$ & ... & ... & $\le13.06$\tm{a}\\ 
\enddata 
\tn{a}{3$\sigma$ upper limits are derived from the r.m.s. noise measured over the $3\sigma$ velocity range on each side of the \OI\ velocity.}
\tn{b}{The $b$-value was held constant to obtain a better fit to the absorption line. For the HVC absorption toward HE\,0027, \OI\ was held to match that of \SiII. For the HVC toward IRAS\,0459, $b$\ion{N}{1} was held to match $b$(\OI). For the HVC toward Mrk\,969, $b$(\OI) was held to match $b$(\CII). For the HVC toward UVQS\,J0110, $b$(\FeII) was derived from \FeII\ $\lambda1608$ alone.}
\end{deluxetable*}

\subsection{\HI\ Observations}

As part of a joint HST-GBT program (GBT project ID HST270191), we obtained frequency-switched single-pointing L-band GBT data for each sight line over a 9\farcm1 beam. We calibrated the data using standard techniques and corrected for stray-radiation using the methods described in \citet{Boothroyd_2011}. We fit third to seventh order polynomials to the emission-free channels to remove residual instrumental baselines. We then smoothed the resulting spectra to a channel width of 1 \kms. 

We fit Gaussians to the \HI\ emission at the velocity of each HVC using custom Python scripts and calculated the column densities ($N_{\mathrm{HI}}$) of each HVC using the standard relation: 
\begin{equation}\label{equation:nhi}
N_{\mathrm{HI}}[\mathrm{cm}^{-2}]=1.823\times10^{18} \int T_{B}(v)~ dv\:[\mathrm{K}\:\mathrm{km}~\mathrm{s}^{-1}]
\end{equation}
\noindent where $T_{B}(v)$ is the velocity-dependent brightness temperature in Kelvin, integrated over the emission velocity range in \kms. Using the Gaussian fits, we calculate $\int T_B(v)~ dv\,[\mathrm{K}\,$\kms] $= 1.064\, h\,\mathrm{FWHM}\,[\mathrm{K}\,$\kms], where $h$ is the height of the fit and FWHM is the full width at half maximum. We then calculate the $b$-value using FWHM=1.665$b$. The results of the Gaussian fits to the \HI\ emission can be seen in Table~\ref{table:fits}. A plot showing the \HI\ spectra for each sight line is given in Figure~\ref{figure:cts-uv-fits}.

\section{Results}\label{section:results}

All five HVCs in our sample show detections of strong UV absorption in low- and medium-ionization metal lines, including: \OI\ $\lambda$1302, \CII\ $\lambda$1334, \AlII\ $\lambda$1670, \FeII\ $\lambda\lambda$ 1608, 1144, \SiII\ $\lambda\lambda\lambda$ 1190, 1193, 1260, and \SiIII\ $\lambda$1206. 
Two of the five HVCs, G348.3–83.8–192 and G133.5–75.6–294, (those toward HE\,0027 and Mrk\,969) are detected in the high-ion doublets \CIV\ $\lambda\lambda$ 1548, 1550 and \SiIV\ $\lambda\lambda$ 1393, 1402. 

Of the low-ion absorption lines covered, \OI\ $\lambda1302$ is considered the most useful for interstellar metallicity measurements, because oxygen has low levels of dust depletion \citep{Jensen_2005, Jenkins_2009} and small ionization corrections (ICs) at large \HI\ column densities \citep[log\,$N_{\mathrm{HI}}\gtrsim18.5$;][]{Bordoloi_2017}. Therefore, we use \OI\ $\lambda1302$ for all our HVC metallicity measurements. 

\subsection{Cloudy Models}\label{section:cloudy}

We calculate the ICs by running a customized suite of \textit{Cloudy} photoionization 
models \citep{Ferland_2017}, which simulate the ionization conditions in the HVCs. 
The models assume a plane-parallel geometry with uniform gas density. The slab is illuminated by the position-dependent Galactic radiation field presented in \citet{Fox_2014}, which is based on \citet{Bland_Hawthorn_1999} and \citet{Fox_2005}. We also include the cosmic-ray background \citep{Indriolo_2007} and extragalactic background radiation \citep{Khaire_2019}. Since we do not know the distance to the HVCs, we scale the flux of hydrogen-ionizing photons, $\Phi$(H), to a range of distances from $5-150$ kpc, specifically 5, 10, 20, 50, 75, 100, and 150 kpc. $\Phi$(H) is calculated based on the Galactic coordinates and distance of the cloud using a custom Python script\footnote{\url{https://github.com/Deech08/DK_HST_Tools}} that takes the three-dimensional radiation field from \citet{Bland_Hawthorn_2019} and \citet{Antwi-Danso_2020}. We use the range $5-150$ kpc to cover nearby clouds and HVCs that are at larger distance estimates of the Magellanic Stream \citep[e.g.][]{Besla_2012}. We also run a \textit{Cloudy} model for each HVC without a Galactic radiation field (i.e., UV background only) to represent an extragalactic cloud. These extragalactic models correspond to our sample of HVCs being at distances $\gtrsim$150--200~kpc (depending on the cloud's $l$ and $b$).

\begin{figure}[!ht]
\centering
\epsscale{1.1}
\plotone{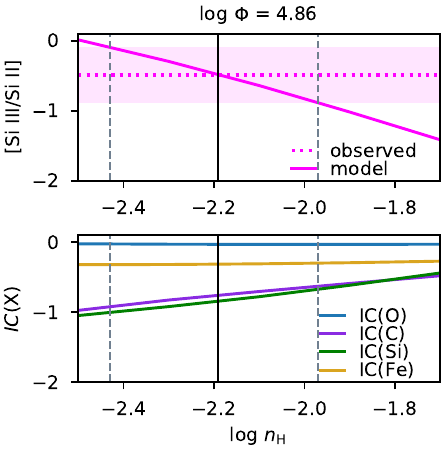}
\caption{Illustration of our \textit{Cloudy} methodology for deriving the gas density, using the example of the HVC toward CTS\,47. 
The top plot shows the modeled (solid line) and measured (dotted pink line) \SiIII/\SiII\ ratio as a function of gas density, with the measurement errors represented by the shaded region. The intersection of the model and measurement gives the value of log\,$n_{\rm H}$ used in the final calculations, denoted by the black dashed vertical line. The corresponding errors on log\,$n_{\rm H}$ are denoted by the grey dashed vertical lines. 
The bottom panel shows the modeled ICs for four elements under study (O, C, Si, and Fe).}
\label{figure:cloudy}
\end{figure}

\begin{figure}[!ht]
    \centering
    \epsscale{1.2}
    \plotone{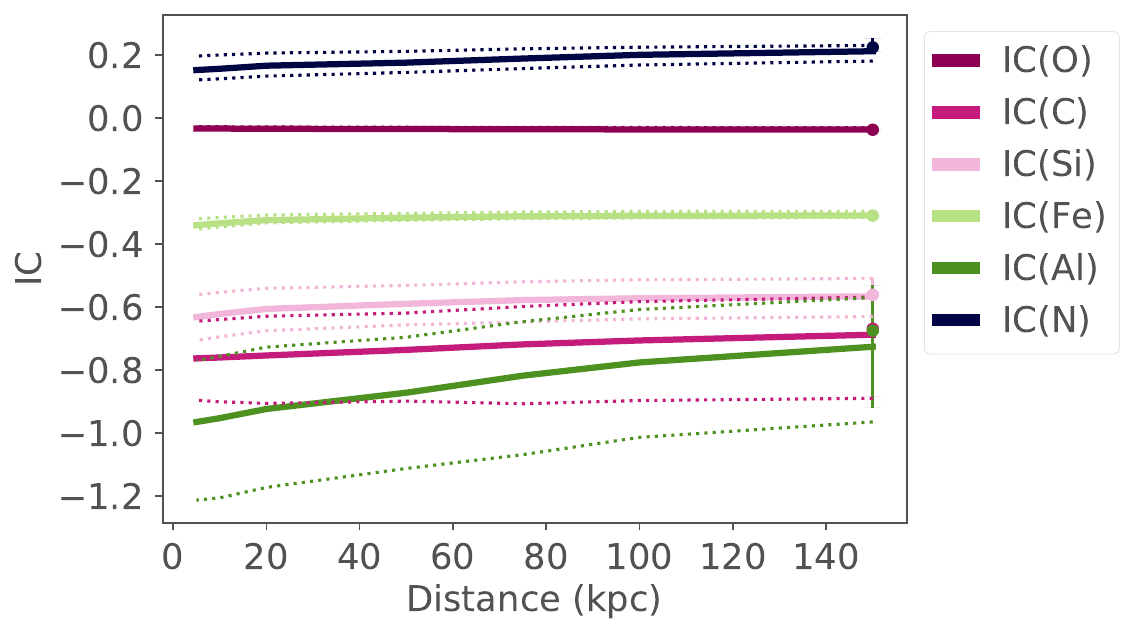}
\caption{Ionization corrections (ICs) from \emph{Cloudy} models over all tested distances for the HVC toward CTS\,47. Solid lines represent the ICs and dotted lines represent their associated errors. Filled circles at 150 kpc represent ICs and their errors (vertical lines) for ``extragalactic" models without a MW UV field.}
\label{figure:IC_dist}
\end{figure}

For each distance, we run \textit{Cloudy} with a grid of hydrogen number densities log $n_\mathrm{H}$ ranging from $-3.0$ to $0.0$ in steps of $0.2$ dex, where $n_\mathrm{H}$ is in cm$^{-3}$.  To constrain $n_\mathrm{H}$, we plot the modeled \SiIII/\SiII\ ratios against $n_\mathrm{H}$ and compare them to the measured \SiIII/\SiII\ value and its $\pm1\sigma$ errors. For more details on the IC error handling, see Appendix~\ref{appendix:cloudy}. An example of the modeled \SiIII/\SiII\ vs. log $n_\mathrm{H}$ and the modeled ICs vs. log $n_\mathrm{H}$ is shown in Figure~\ref{figure:cloudy}, which models the HVC toward CTS\,47. 
We then rerun the \textit{Cloudy} models with the exact interpolated $n_\mathrm{H}$ values that match the measured \SiIII/\SiII\ and the associated $\pm1\sigma$ errors. The ICs are then calculated for each distance and added to the abundance of that ion. In Figure~\ref{figure:IC_dist} we show an example of the IC dependence on distance, again using the example of the cloud toward CTS\,47. Importantly, we find that the modeled ICs do not depend strongly on distance, with the exception of aluminum, whose IC becomes slightly less negative with increasing distance.

\subsection{Abundances and Depletion}\label{section:abundances}

We calculate the gas-phase abundance of element X as:

\begin{equation}\label{equation:XiH}
\left[\frac{\rm{X}}{\rm{H}}\right]=\mathrm{log}\left(\frac{N_{\mathrm{X}^i}}{N_\mathrm{H\;I}}\right) - \mathrm{log}\left(\frac{\rm{X}}{\rm{H}}\right)_{\odot}+{\rm IC}({\rm X}), 
\end{equation}

\noindent where X$^i$ is the observed ion, $N_{\mathrm{X}^i}$ and $N_\mathrm{H\;I}$ are the ionic and atomic hydrogen column densities, respectively, and log\,(X/H)$_\odot$ is the elemental solar abundance taken from \citet{Asplund_2009}. When calculating the oxygen-based metallicity, we add a beam-smearing error to the hydrogen columns of 0.11 dex. This 1$\sigma$ error is determined by comparing low- and high-resolution \HI\ maps of HVCs, as discussed in detail in Appendix~\ref{appendix:beam_smearing}. We present the HVC oxygen-based metallicity measurements in Table~\ref{table:metallicities}. 

\begin{deluxetable*}{lcccccc}[!ht]
\tabletypesize{\small}
\tablecaption{HVC Ionization Corrections and Metallicity Measurements}\label{table:metallicities}
\tablehead{\colhead{Sight Line}  & \colhead{$v_{0\: \mathrm{UV}}$} & \colhead{X$^{i}$}   & \colhead{[X$^{i}$/H]\tm{\footnotesize a}}   & \colhead{IC\tm{\footnotesize b}} & \colhead{[X/H]\tm{\footnotesize c}} & \colhead{$Z$\tm{\footnotesize c}} \\[-6pt]
 \colhead{} & \colhead{(\kms)} & \colhead{}  &  \colhead{}  & \colhead{($5$ kpc : Extragalactic)} & \colhead{($5$ kpc : Extragalactic)}  & \colhead{($Z_\odot$)}   }
\startdata
CTS\,47    & $155.9\pm7.6$ & \OI\ & $-1.32\pm0.13$ & $-0.036:-0.032$ & $-1.36:-1.36$ & $0.044\pm0.014$\\
HE\,0027    & $-166.9\pm8.9$  &  \OI\ & $-0.77\pm0.16$ & $-0.038:-0.028$ & $-0.79:-0.80$ & $0.16\pm0.06$\\
IRAS\,0459   & $130.8\pm8.1$  &  \OI\ & $-0.77\pm0.15$ & $-0.008:-0.003$ & $-0.77:-0.78$ & $0.17\pm0.06$\\
Mrk\,969  & $-305.8\pm11.2$ & \OI\  & $-1.66\pm0.15$ & $-0.028:-0.027$ & $-1.69:-1.69$ & $0.021\pm0.002$\\
UVQS\,J0110  & $290.4\pm7.8$ &  \OI\ & $-1.70\pm0.13$ & $-0.030:-0.016$ & $-1.72:-1.73$ & $0.019\pm0.006$ \\
\enddata
\tn{a}{Gas-phase ion abundance, [X$^i$/H]=[log\,$N$(X$^i$)--log\,$N$(\HI)]--log\,(X/H)$_{\sun}$. We use a solar oxygen abundance of $10^{-3.31}$ \citep{Asplund_2009}. Errors are added in quadrature and include a 0.11 dex error estimate for beam smearing. See Appendix~\ref{appendix:beam_smearing} for details on the calculation of this error.}
\tn{b}{Range of IC values calculated for a range of distances that the HVCs might plausibly cover.}
\tn{c}{Metallicities are presented in both logarithmic and linear forms.}
\end{deluxetable*}

We plot the oxygen, sulfur, carbon, silicon, iron, aluminum, and nitrogen, gas-phase abundances measured in each sight line at each distance in Figure~\ref{figure:abundance} (values are listed in Appendix~\ref{appendix:cloudy}). The HVCs each have sub-solar gas-phase abundances of each element (uncorrected for dust) and follow a fairly similar abundance pattern, with an average difference between the measured HVC abundances of \s1.0 dex. 
The two HVCs showing higher oxygen-based metallicity (the clouds toward HE\,0027 and IRAS\,0459) consistently have higher abundance measurements than the rest of the HVCs in all elements detected. 
For oxygen, the HVCs toward HE\,0027 and IRAS\,0459 have nearly the same abundance, while for most other elements the HE\,0027 cloud has a higher abundance than the IRAS\,0459 cloud, which could be an indication that the HE\,0027 cloud has less dust depletion than the IRAS\,0459 cloud in those elements.  

\begin{figure*}[!ht]
    \centering
    \epsscale{1.18}
      \plotone{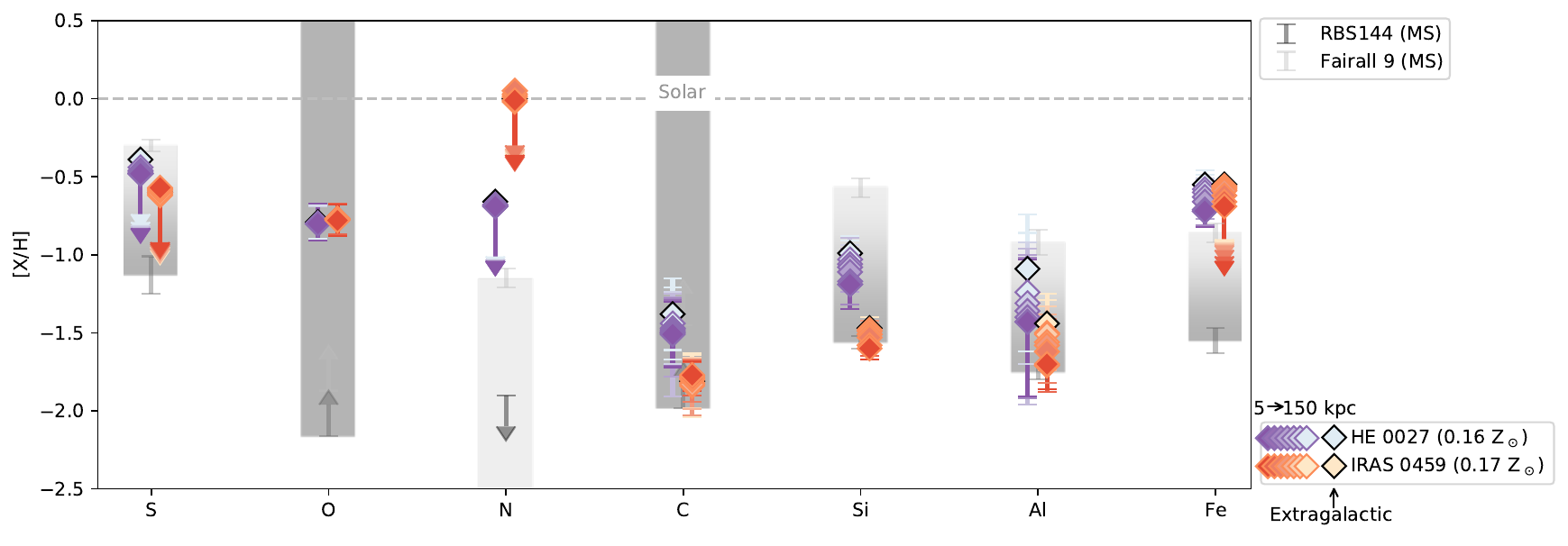}
    \epsscale{1.2}
      \plotone{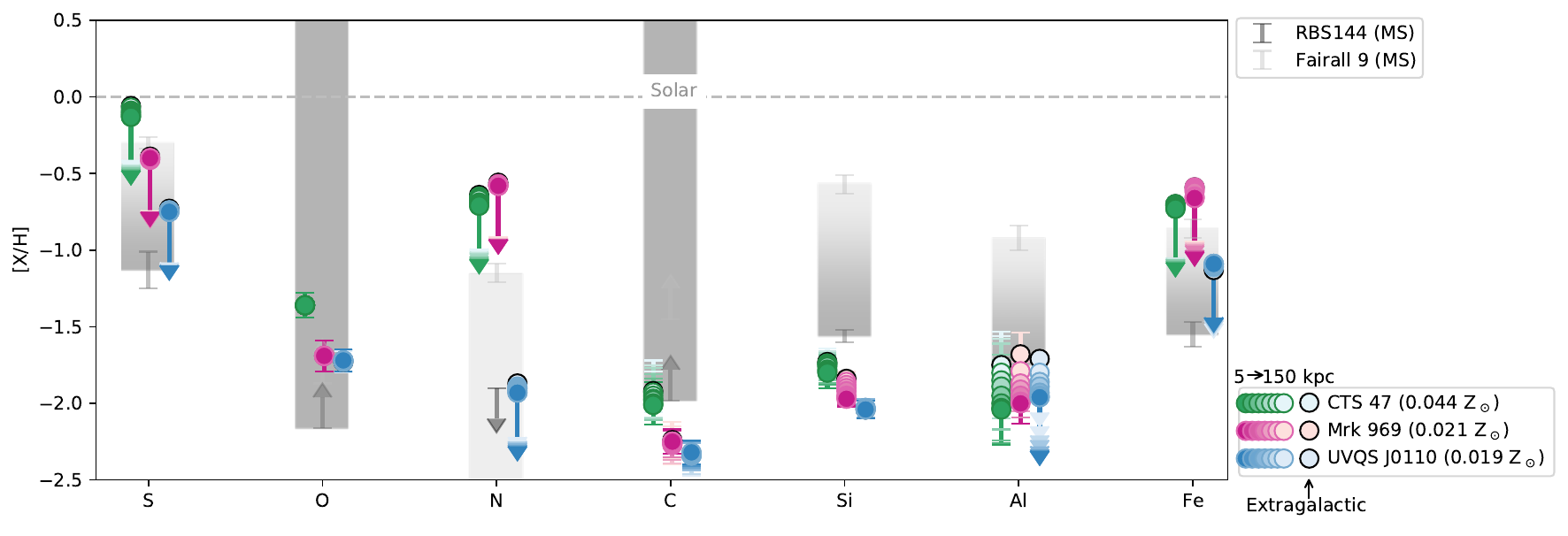}
\caption{Gas-phase abundances measured for the high-metallicity (top panel) and low-metallicity (bottom panel) HVCs. Comparison data on the Magellanic Stream are shown as grey shaded regions with arrows indicating limits \citep[from the RBS\,144 and Fairall\,9 sightlines; dark grey and light grey, respectively;][]{Fox_2013, Richter_2013}.}
\label{figure:abundance_comparison}
\end{figure*}

In Figure~\ref{figure:abundance_comparison} we have separated the abundances into two plots: low- and high-metallicity clouds, and have compared our HVC sample abundance measurements to those of the Magellanic Stream \citep[MS;][]{Richter_2013, Fox_2013}. 
The high-metallicity HVCs closely follow the abundance patterns of the MS. 
The low-metallicity clouds (towards CTS 47, Mrk 969, and UVQS J0110) only partially fall within the MS abundance pattern. Both their carbon (with the exception of CTS 47) and silicon gas abundances are outside the bounds of the MS abundances. 

\begin{figure*}[!ht]
    \centering
    \epsscale{1.15}
      \plotone{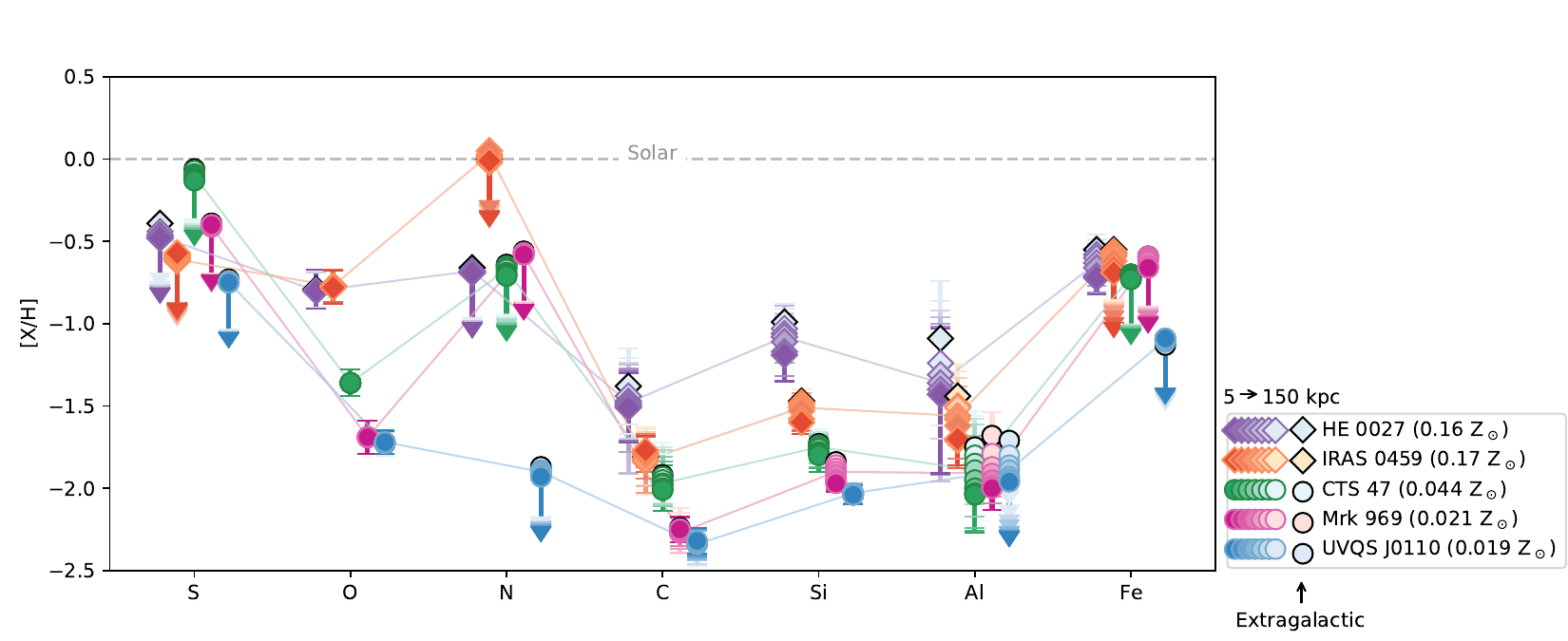}
\caption{Gas-phase abundances measured in each HVC; measurements for low-metallicity and high-metallicity sight lines are denoted by circles and diamonds, respectively, with error bars. Upper limits are indicated by an arrow. The measurements are corrected for ionization using \textit{Cloudy} models at each distance (5--150 kpc) and one with no MW UV radiation field (``Extragalactic"). The gradient in colors represent different distances for the HVCs with darker colors representing smaller distances. The extragalactic model is denoted by a symbol with a dark outline. The grey-dashed horizontal line indicates solar abundance. For each sight line, we connect the various elements with a solid line as a visual aid to show the abundance pattern.}
\label{figure:abundance}
\end{figure*}

For each distance, we calculated the dust depletion of each element relative to oxygen, since oxygen is taken to be undepleted onto dust grains: 

\begin{equation}\label{equation:dust}
        \delta({\rm X})_{\rm O} = \left[\frac{\rm X}{\rm H}\right] - \left[\frac{\rm O}{\rm H}\right]. 
\end{equation}

\noindent The dust depletion values for each HVC for each assumed distance are listed in Appendix~\ref{appendix:dust} and plotted in Figure~\ref{figure:oxygen_depl}. 
The HVCs all show a similar depletion pattern, with significant depletion seen in carbon, silicon, and aluminum relative to oxygen. Such depletions 
are usually taken as evidence for dust, but they could also indicate an intrinsically non-solar abundance pattern, as the HVCs may arise 
outside of the Galactic ISM and therefore have a very different nucleosynthetic history than standard ISM gas.

Only the HE\,0027 HVC has an iron measurement (showing very little depletion in iron), while the four other HVCs only show upper limits on iron.
The HVCs do not follow a clear pattern of depletion depending on metallicity. 
The HVCs generally have similar depletion levels to the MS in carbon, silicon, and aluminum 
\citep[using the MS depletion patterns from][]{Richter_2013, Fox_2013}.
However, the match is not exact; for example, IRAS 0459 and HE 0027 have MS-like metallicities, but do not match the MS depletions of silicon and iron, respectively.

\begin{figure*}[!ht]
    \centering
    \epsscale{1.2}
    \plotone{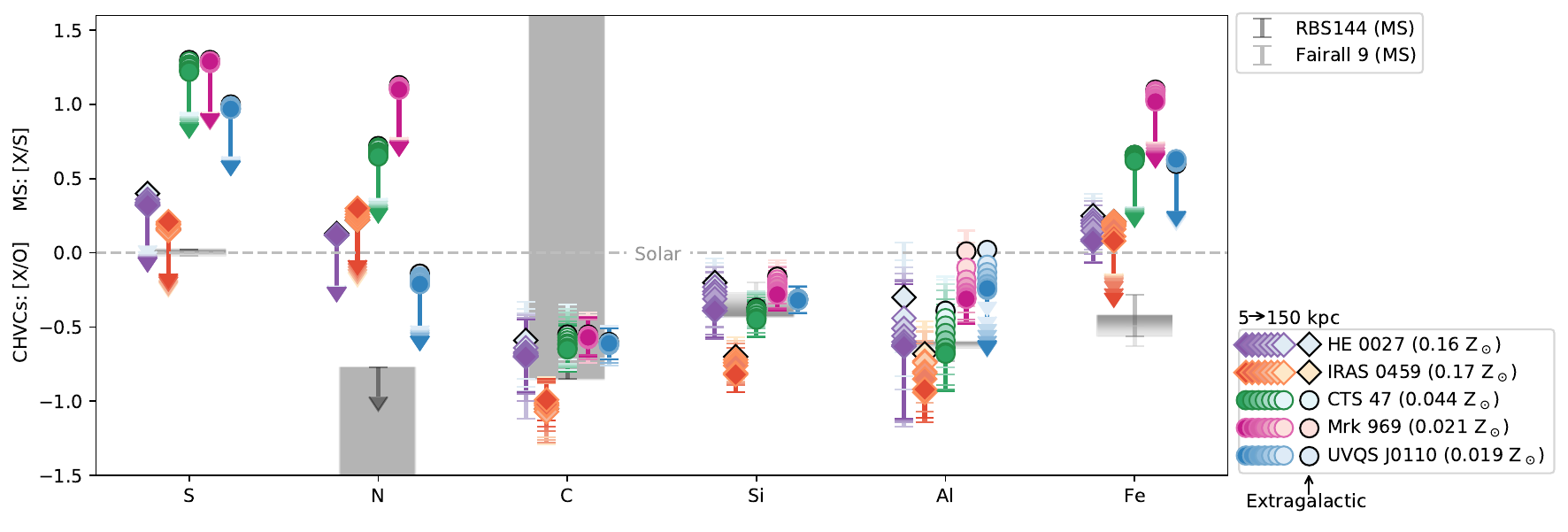}
    \caption{Depletion relative to oxygen and sulfur in each HVC, compared to the Magellanic Stream depletion pattern. All symbols are the same as in Figure~\ref{figure:abundance_comparison}.}
\label{figure:oxygen_depl}
\end{figure*} 

\section{Discussion}\label{section:discussion}

The origins of compact and small HVCs are poorly understood, and are likely to be diverse. They could be nearby clouds in the Galactic fountain, fragments of the Magellanic System, dwarf galaxies with low star formation rates in the Local Group, minihalos, or clouds in the intergalactic medium \citep{Oort_1966, Giovanelli_1978, Bregman_1980, Braun_1999, Blitz_1999, Bruns_2001, Sternberg_2002, deHeij_2002c, Westmeier_2005,Pisano_2004, Pisano_2007, Putman_2006, Putman_2012}. 

Metallicities provide key information to constrain the origin of these clouds,
because different origins will lead to different metallicities.
Disk clouds and fountain clouds are expected to be metal-enriched, with \s$0.50$ solar to super-solar metallicity, whereas halo clouds, such as the well-studied HVC Complex C, have lower metallicities of $0.10-0.30$ $Z_\sun$ \citep{Wakker_1999, Gibson_2000, Richter_2001, Tripp_2003, Fox_2023}. Magellanic Stream clouds typically have metallicities of $\approx0.10$ $Z_\sun$ \citep{Fox_2010, Fox_2013}, but can reach $\approx0.50$ $Z_\sun$ in certain directions \citep{Richter_2013}. Local Group dwarf irregular galaxies have metallicities of \s$0.03-0.30$ $Z_\sun$ \citep{Kunth_2000}. The low-redshift intergalactic medium (IGM) has been modeled and measured to have a metallicity of \s$0.10$ $Z_\sun$ \citep{Ferrara_2000, Tripp_2002, Shull_2003}. 

Our new measurements show that all five small HVCs in our sample have a low metallicity of $\le0.17$\,$Z_\sun$, and that three of the five have very low metallicities of $\le0.04$\,$Z_\sun$ (see Table~\ref{table:metallicities}). All of these metallicities are oxygen-based, and so are robust against ionization and dust corrections. We have also performed a thorough analysis of beam-smearing in HVCs and have included the results in our metallicity errors (see Appendix~\ref{appendix:beam_smearing}). These low metallicities indicate that the small HVCs are not ejected MW disk gas and instead originate further away, either in the Galactic halo, the Magellanic System, or the Local Group.

All five small HVCs are projected close to the Magellanic System (see Figure~\ref{figure:westmeier}), so one simple explanation is that they are all fragments of the Magellanic System that have been detached by tidal or hydrodynamic interactions. This explanation has been favored for other small HVCs in the past,
for example the Compact HVC toward the QSO Ton S210 \citep{Sembach_2002, Kumari_2015}, a sightline that passes close to the Magellanic System.
\citet{Westmeier_2008} and \citet{Nidever_2010} suggest that the HVCs towards Mrk\,969 and UVQS\,J0110 are likely part of the MS based on their \HI\ position and velocities alone.
For comparison, the SMC and LMC have current-day metallicities $0.22$ $Z_\sun$ and $0.46$ $Z_\sun$ ([O/H]$_{\rm SMC}= -0.66\pm0.10$ and [O/H]$_{\rm LMC}= -0.34\pm0.06$; \citealt{Russell_1992, Asplund_2009}). The Magellanic Bridge and large parts of the Magellanic Stream have lower metallicities of $0.11$ and $0.10$ $Z_\sun$, respectively, based on the measurements [O/H]$_{\rm MB} = -0.96\pm^{0.13}_{0.11}$ \citep{Lehner_2002} and [O/H]$_{\rm MS}$ = $-1.00\pm0.09$ \citep{Fox_2013}, though higher metallicities of up to \s$0.50$ $Z_\sun$ have been measured in the LMC filament of the Stream \citep{Richter_2013}. 

The HVCs towards HE\,0027 and IRAS\,0459 have metallicities of $0.16\pm0.07$ and $0.17\pm0.07$ $Z_\sun$, respectively, similar to the values seen in the SMC and MS. They also have similar relative abundance patterns to that of the MS (see Section~\ref{section:abundances}). The HVC towards HE\,0027 (HVC G348.3–83.8–19) appears to be projected between two filament in the MS and is at a similar velocity to the surrounding MS gas. Therefore, its velocity, projected location, and abundance pattern indicate that HVC G348.3–83.8–19 is very likely a MS fragment. The HVC towards IRAS\,0459 (HVC G224.0–34.3+135) is projected \s40\degr\ away from the edge of the LMC, but may still be Magellanic given that the Leading Arm and Stream combined reach a total of \s210\degr\ in length \citep{Nidever_2010} and show ionized gas absorption up to 30\degr\ away from the \HI\ \citep{Fox_2014}. Additionally, there is a group of small clouds centered on a Magellanic longitude of \s15\degr\ and Magellanic latitude of \s30\degr\ of similar velocities (\s100\,\kms; see Figure~\ref{figure:westmeier}) that could connect HVC G224.0–34.3+135 to the rest of the Magellanic System. Therefore, we conclude that HVC G224.0–34.3+135 towards IRAS\,0459 is likely a fragment of the Magellanic System.

The other three HVCs in our sample, toward CTS\,47, Mrk\,969, and UVQS\,J0110, have much lower metallicities of $0.02-0.04$ $Z_\sun$, which are significantly lower than those measured in the present-day Magellanic System. The cloud metallicities appear too low to explain with a Magellanic origin even when accounting for the possibility of abundance variations across the Magellanic Clouds, because no radial abundance gradient is seen in \ion{H}{2}-region metallicities in either the LMC or SMC \citep{toribio2017}, and only shallow stellar abundance gradients are seen \citep{Cioni_2009, feast2010}, although we cannot rule out the possibility that the clouds had a steeper abundance gradient before they were stripped.
Even in the presence of strong beam-smearing, only one of the three low-metallicity HVCs (G237.2–41.1+146 towards CTS\,47) has a metallicity close to that of the MS, with $Z+1\sigma\approx0.10\,Z_\sun$ (see Appendix~\ref{appendix:beam_smearing}). 

For these low-metallicity HVCs to originate in the Magellanic System, they would have to have formed early in its history (several Gyr ago), when its chemical abundances were much lower. SMC metallicity histories modeled by \citet{Pagel_1998} and \citet{Cignoni_2013} indicate that the SMC had a metallicity similar to that of the HVCs toward Mrk\,969 (HVC G133.5–75.6–29) and UVQS\,J0110 (HVC G146.2–77.6–279) \s$10$ Gyr ago. Similarly, the HVC towards CTS\,47 (HVC G237.2–41.1+146) has a similar metallicity to the SMC $7.5$ Gyr ago in the models of \citet{Tsujimoto_2009} or \s$9-14$ Gyr ago in the early formation history of the galaxies in the models of \citet{Pagel_1998, Cignoni_2013}. 
However, in this scenario the unenriched clouds would need to survive over many Gyr without being
destroyed or enriched while the rest of the Magellanic System increased in metallicity. 
This seems unlikely.

Metal mixing could potentially be invoked as an explanation for the very low metallicities. 
In a metal-mixing scenario, cool clouds can exchange material with the surrounding hot plasma, resulting in a cloud with a metallicity in-between its initial metallicity and that of the surrounding hot halo \citep{Gritton_2014, Gronke_2018,heitsch2022}. However, for a cloud to be mixed down from initial metallicities of $0.10-0.50$ $Z_\sun$ to metallicities of $0.02-0.04$ $Z_\sun$, the surrounding plasma (the diluting medium) would itself have to have a metallicity of $\le0.02-0.04$ $Z_\sun$. Even in the recently discovered Magellanic Corona, \citet{Krishnarao_2022} estimate the metallicity to be ${\rm [Z/H]}\approx-1$ or $0.10$ $Z_\sun$, higher than the values in HVCs G133.5–75.6–29, G146.2–77.6–279, and G237.2–41.1+146. It is therefore difficult to argue for a Magellanic origin for these three HVCs, even after accounting for metal mixing.

An alternative explanation for the small HVCs is that they trace intergalactic gas in the Local Group, beyond the Magellanic Clouds. However, it is unclear whether IGM clouds would have such high \HI\ column densities. 
Most IGM gas is expected to be in the warm and hot phase due to the ionizing UV and X-ray background, with only $<1$\% in a cold neutral form and most of the \HI\ reaching columns of only \s$10^{17}$ cm$^{-2}$ 
\citep{Popping_2011, Lockman_2012, Takeuchi_2014, Kooistra_2017}, 
whereas our sample reaches columns of $>10^{18}$ cm$^{-2}$. 
Additionally, most of the low-redshift IGM is expected to be enriched to at least \s$0.10$ $Z_\sun$ by metal ejection from galaxies \citep[with the addition of small amounts of enrichment from Pop III stars;][]{Ferrara_2000, Tripp_2002, Shull_2003}, which is $2-5$ times higher than the $0.02-0.04$ $Z_\sun$ measured in the HVCs G133.5–75.6–29, G146.2–77.6–279, and G237.2–41.1+146. For all these reasons, we conclude that the HVCs are not likely to trace the IGM.

Another potential explanation for the three low-metallicity HVCs (G133.5–75.6–29, G146.2–77.6–279, and G237.2–41.1+146) is that they are related to dwarf galaxies. Two possible scenarios are that the clouds \emph{are} ultra-faint dwarf galaxies whose stars are currently undetected, or are \emph{removed} from dwarf galaxies by tidal interactions or ram-pressure stripping \citep{Braun_1999, Blitz_1999, deHeij_2002c}. 
Typical \HI\ sizes of dwarf galaxies are $\approx$2--10 kpc \citep{Hunter_2012}, so our small HVCs would
have to be at distances of $\approx$60--300 kpc to be dwarf galaxies, which would place them well within the Local Group, inside the Milky Way's virial radius.
Extremely metal-poor galaxies (XMPs; $<0.10$ $Z_\sun$) are not uncommon, making up \s20\%\ of faint blue dwarf galaxies \citep{James_2015, James_2016}. Several diffuse dwarf galaxies have been discovered with metallicities $<$0.03\,$Z_\sun$, including Leo P, AGC 198691, Little Cub, and J0811+4730 \citep{Giovanelli_2013, Skillman_2013, Hirschauer_2016, Hsyu_2017, Izotov_2017}. Dwarf galaxies can also show considerable metallicity gradients, 
so that gas on their outskirts can have very low metallicities \citep{Taibi_2022}.
They also often display structure in their \HI\ content \citep{deblok1996, begum2006}, which could potentially explain the structure we observe in our HVCs (Figure~\ref{figure:chvc_hi}).

The key challenge for the dwarf-galaxy explanation is the lack of stellar counterparts. We visually inspected DSS, WISE, and 2MASS images of the HVC fields to search for faint stellar components, but none were found. We also conducted a nearby galaxy search within 5\arcmin\ of each HVC using the NASA/IPAC Extragalactic Database (NED), and did not find any known galaxies. This points away from a dwarf-galaxy origin, although it is possible that enough time may have passed for the stellar component to be separated from the gas by an appreciable distance, especially in the presence of a massive host galaxy \citep{Blitz_2000, Pearson_2016}. To find an optical counterpart in this case, kinematically-similar optical counterparts would need to be identified and models of the three dimensional kinematics for both nearby stellar components and the HVC would need to be conducted, which is beyond the scope of this paper. 

Finally, the three low-metallicity HVCs could be gaseous minihalos. Minihalos are small dark-matter halos 
with no current star formation that accumulate gas in the early universe and are potentially the building 
blocks of larger galaxies such as the MW \citep{Rees_1986}. Minihalos were first suggested as an explanation 
for HVCs by \citet{Braun_1999} and \citet{Blitz_1999}; they would naturally lack a stellar counterpart, 
which has not been found for the three low-metallicity HVCs in our sample. 
Minihalos can also maintain low metallicities assuming they virialize at high redshift 
\citep{Wyithe_2007, Cen_2008},
consistent with the three sight lines that have low-metallicities of $\le$0.04 $Z_\sun$. 
If the small HVCs are minihalos, then they must have distances of $\le$750 kpc, otherwise they would 
be under-pressured, too large (\s10 kpc), and have no analogs in nearby galaxy groups 
\citep{Pisano_2004, Pisano_2007, Sternberg_2002}. Starless HVC analogs are not detected in nearby 
galaxy groups \citep{Pisano_2004, Pisano_2007}. Therefore, assuming the Local Group is not unique 
in containing large starless gas-rich dark matter halos, \citet{Pisano_2007} find that HVCs must 
be within distances of \s90 kpc. Similarly, \citet{Sternberg_2002} favor the minihalos being 
circumgalactic objects with distances of $\le$150 kpc assuming they have similar dark matter distributions to dwarf galaxies.  

However, if the HVCs in our sample are minihalos close to the MW, then it is unclear how they would retain such low-metallicities, because cloud/corona interactions should have destroyed or disrupted the clouds.
Surviving clouds will mix with the hot plasma and end up with a metallicity between that of the original gas cloud and the hot plasma on the timescale of tens of Myr \citep{Gritton_2014, Gronke_2018, Fielding_2022}.  
Lower estimates of the hot-halo metallicity are $\ge$0.30 $Z_\sun$, while higher estimates place it at closer to $\ge$0.60 $Z_\sun$ \citep{Miller_2015, Miller_2016, Gritton_2014, Henley_2017}, both of which would be expected to raise the metallicity of the HVCs well beyond their $\le$0.04 $Z_\sun$ measurements given tens of Myr. Therefore, we conclude that the three low-metallicity HVCs could be minihalos only if they are located outside the MW hot halo or entered it fairly recently.

\section{Summary}\label{section:summary}

We have analyzed HST UV absorption-line spectra and GBT \HI\ emission spectra of five small HVCs chosen from GASS in the southern Galactic sky \citep{McClure-Griffiths_2009,Moss_2013,Moss_2017}. 
The clouds cover between 1 and 2.5$\degr$ on the sky.
Each of the five HVCs has a UV-bright AGN projected behind their observed \HI\ emission within $1$\degr\ of the peak GASS emission. We used these five AGN sight lines together with \textit{Cloudy} photoionization modeling to determine the ionization-corrected gas-phase abundances of each HVC. Our results are as follows: 

\begin{enumerate}
    \item All five of the HVCs have oxygen-based metallicities $<$0.17 $Z_\sun$.
    These low metallicities preclude a MW disk origin for these HVCs.
    
    \item The HVCs toward HE\,0027 (HVC G348.3–83.8–192) and IRAS\,0459 (HVC G224.0–34.3+135) have metallicities of 0.16--0.17 $Z_\sun$, similar to the SMC and the Magellanic Stream. Given that their metallicities, velocities, and projected locations are all consistent with the Magellanic Stream, these two HVCs are likely fragments of the Magellanic System.  
    \item The HVCs toward CTS 47 (HVC G237.2–41.1+146), Mrk 969 (HVC G133.5–75.6–294), and UVQS J0110 (HVC G146.2–77.6–279) all have very low metallicities of $\le$0.04 $Z_\sun$. Despite their similar velocities and nearby projected locations to the Magellanic System, we conclude they are not likely part of the Magellanic System since their metallicity is too low. 
    A plausible alternative explanation for these clouds is that they 
    are stripped from dwarf galaxies by tidal interactions or ram-pressure stripping. 
    Alternatively, the low-metallicity HVCs could be gaseous minihalos without star formation, 
    explaining both the low-metallicity of the HVCs and lack of a stellar counterpart. If they are minihalos, then in order to maintain a low metallicity, they would likely reside outside of the MW hot halo. 
\end{enumerate}

These metallicity results indicate that two of our HVCs are likely to be fragments of the Magellanic System, while the three low-metallicity HVCs could be extragalactic sources such as 
starless minihalos or gaseous debris stripped by tidal interactions or ram pressure. Deep optical or infrared observations towards the three low-metallicity HVCs are needed to search for optical counterparts and further narrow the possible explanations for their gaseous properties. Our study concludes that small HVCs, as a class, have a variety of origins, from Magellanic System fragments to objects in the Local Group.\\

{\it Acknowledgments}: We would like to thank S. Faridani for providing us with the HVC \HI\ data cubes used to estimate beam-smearing. 
We thank the referee for a helpful report.
Support for program 15887 was provided by NASA through a grant from the Space Telescope Science Institute, which is operated by the Association of Universities for Research in Astronomy, Inc., under NASA contract NAS5-26555.

\facilities{HST (COS), GBT, GASS.}  

\software{VPFIT v12.2 \citep{Carswell_2014}, \textit{Cloudy} v17.02 \citep{Ferland_2017}, Python v3.8}.\\

All the {\it HST} data used in this paper can be found in MAST at \dataset[10.17909/26an-rq05]{https://archive.stsci.edu/doi/resolve/resolve.html?doi=10.17909/26an-rq05}.

\clearpage
\bibliographystyle{aasjournal}


\appendix
\restartappendixnumbering

\section{UV absorption}\label{appendix:UV_absorption}

In Figure~\ref{figure:uv-fits-appen} we present the UV absorption profiles and their Voigt-profile fits for each HVC in the sample. 

\begin{figure}[!ht]
    \centering
    \subfloat[CTS 47]{\includegraphics[width=0.9\textwidth]{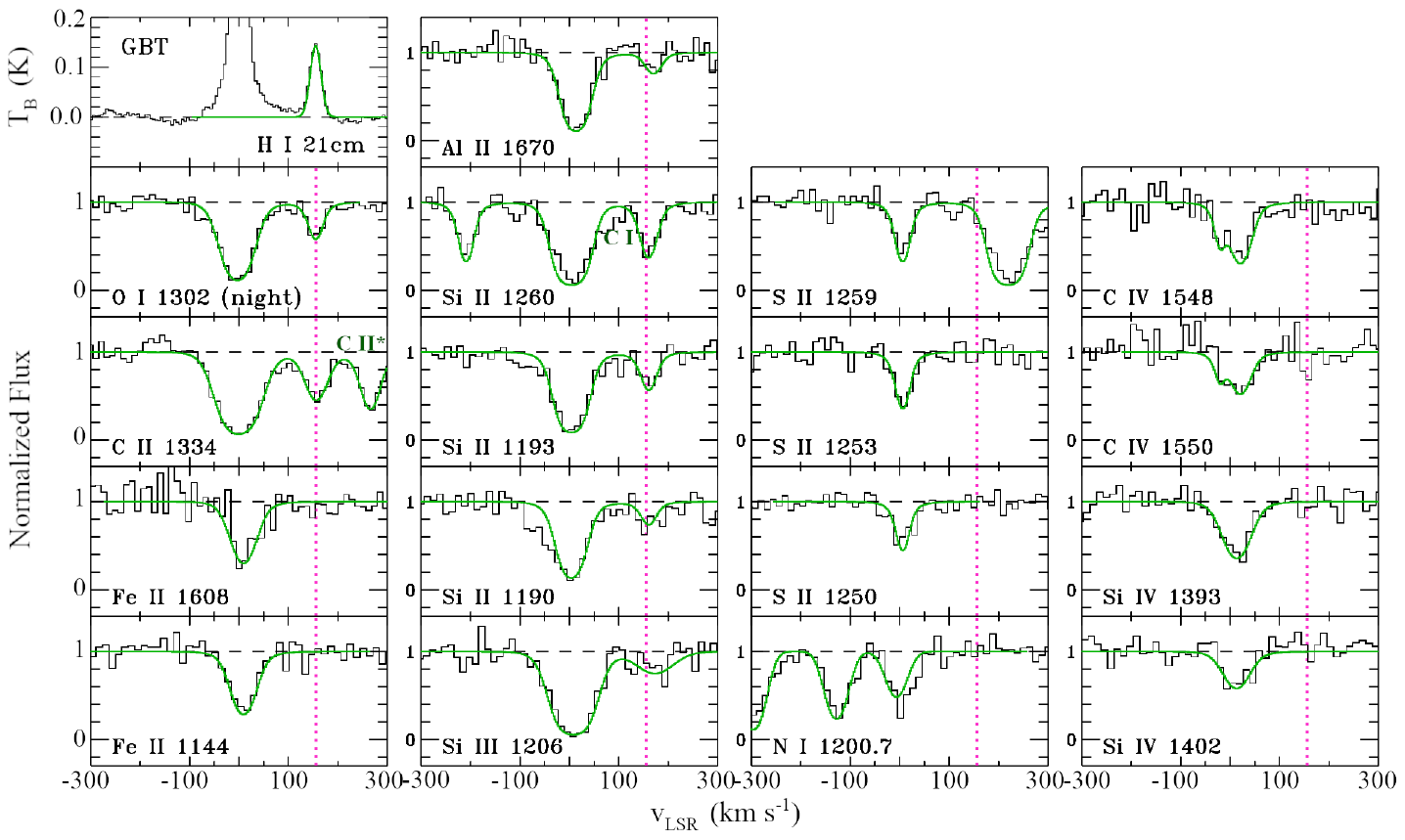}}\\
     \subfloat[HE 0027-3118]{\includegraphics[width=0.9\textwidth]{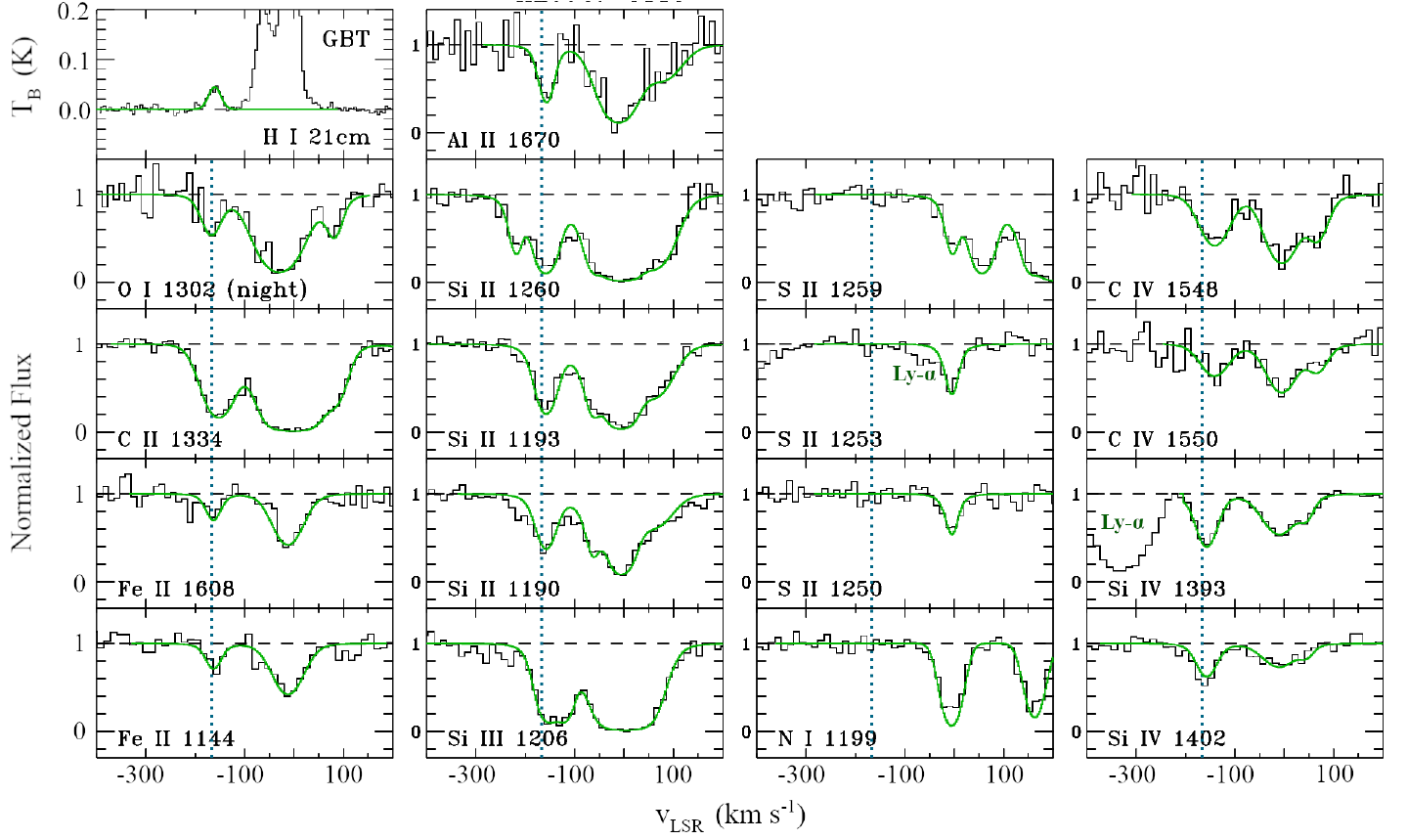}}\\
     \caption{UV absorption-line spectra for metal ions and GBT \HI\ emission spectra for each AGN direction in our sample. The Voigt-profiles fits to each ion and the Gaussian fit to the HVC \HI\ emission are shown in green. The vertical dotted lines represent the velocity centroid of the \OI\  absorption used for metallicity measurements. The \HI\ emission that peaks near \s250 \kms\ in  IRAS 0459 is an artifact in the GBT data and not real.  Mrk 969's \OI\ absorption shows all photons.} \label{figure:uv-fits-appen}
\end{figure}
\begin{figure}
     \ContinuedFloat
     \subfloat[IRAS 04596-2257]{\includegraphics[width=0.9\textwidth]{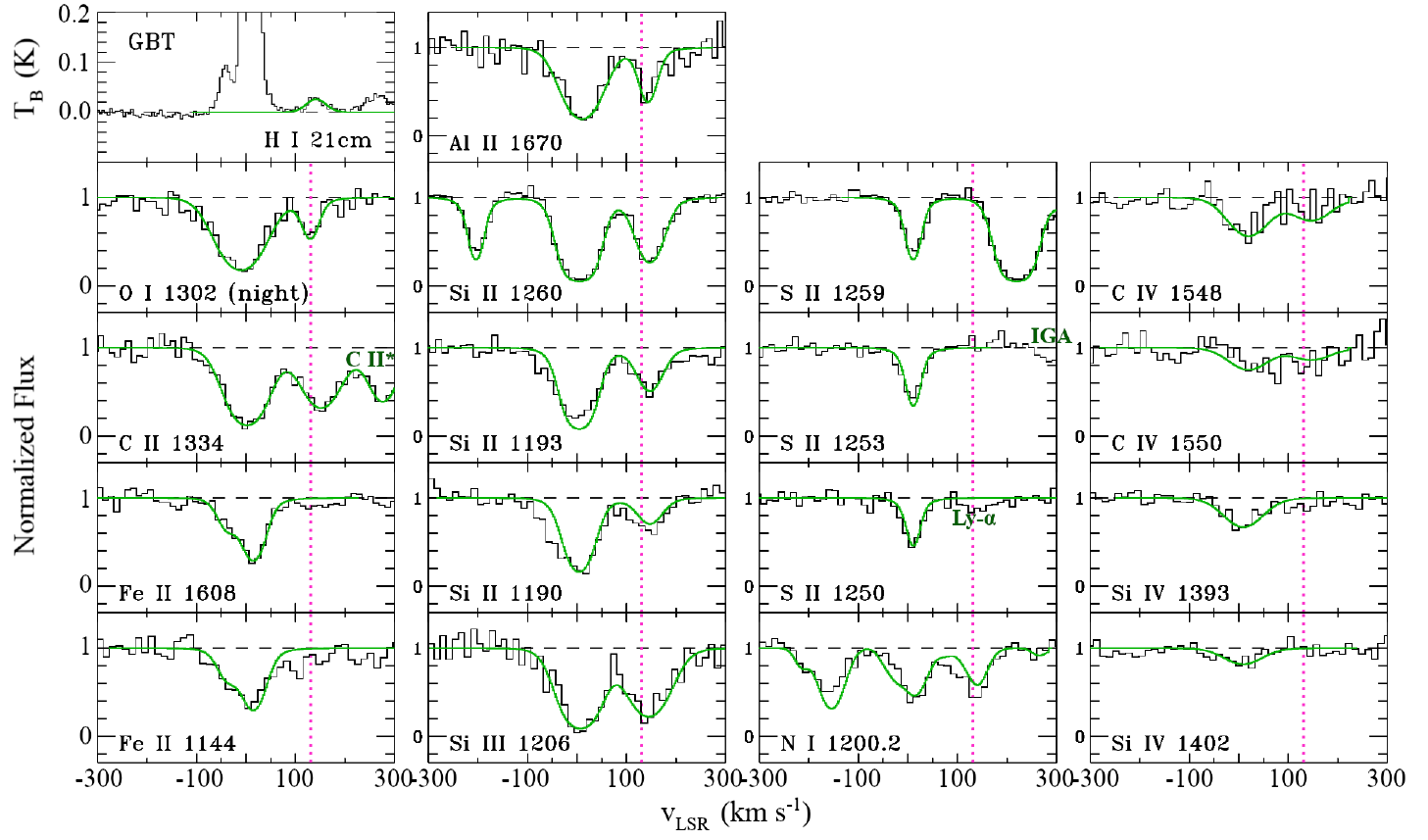}}\\
     \subfloat[Mrk 969]{\includegraphics[width=0.9\textwidth]{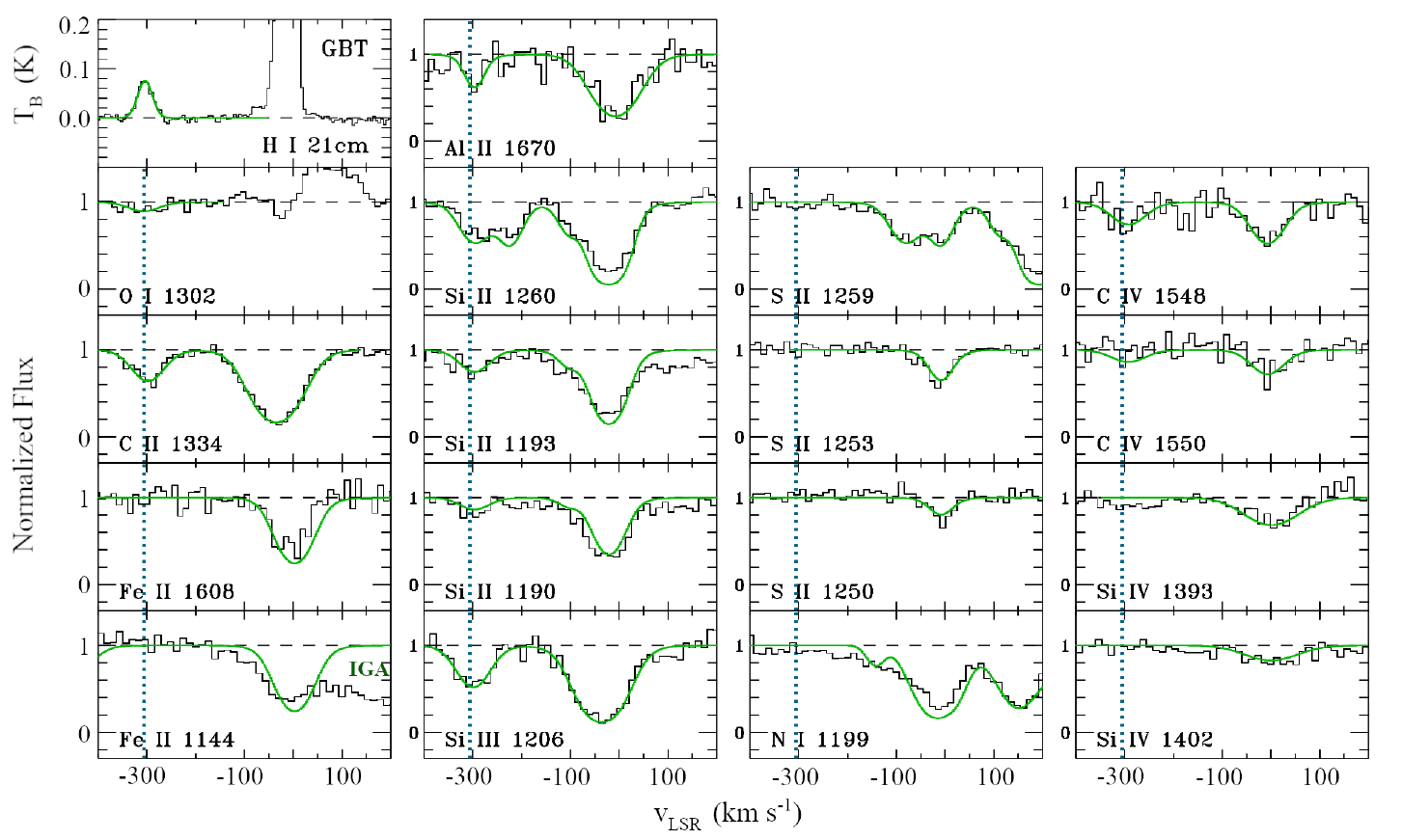}}\\
     \caption{Continued from above.} 
\end{figure}
\begin{figure}
     \ContinuedFloat
     \subfloat[UVQS J011054-154540]{\includegraphics[width=0.9\textwidth]{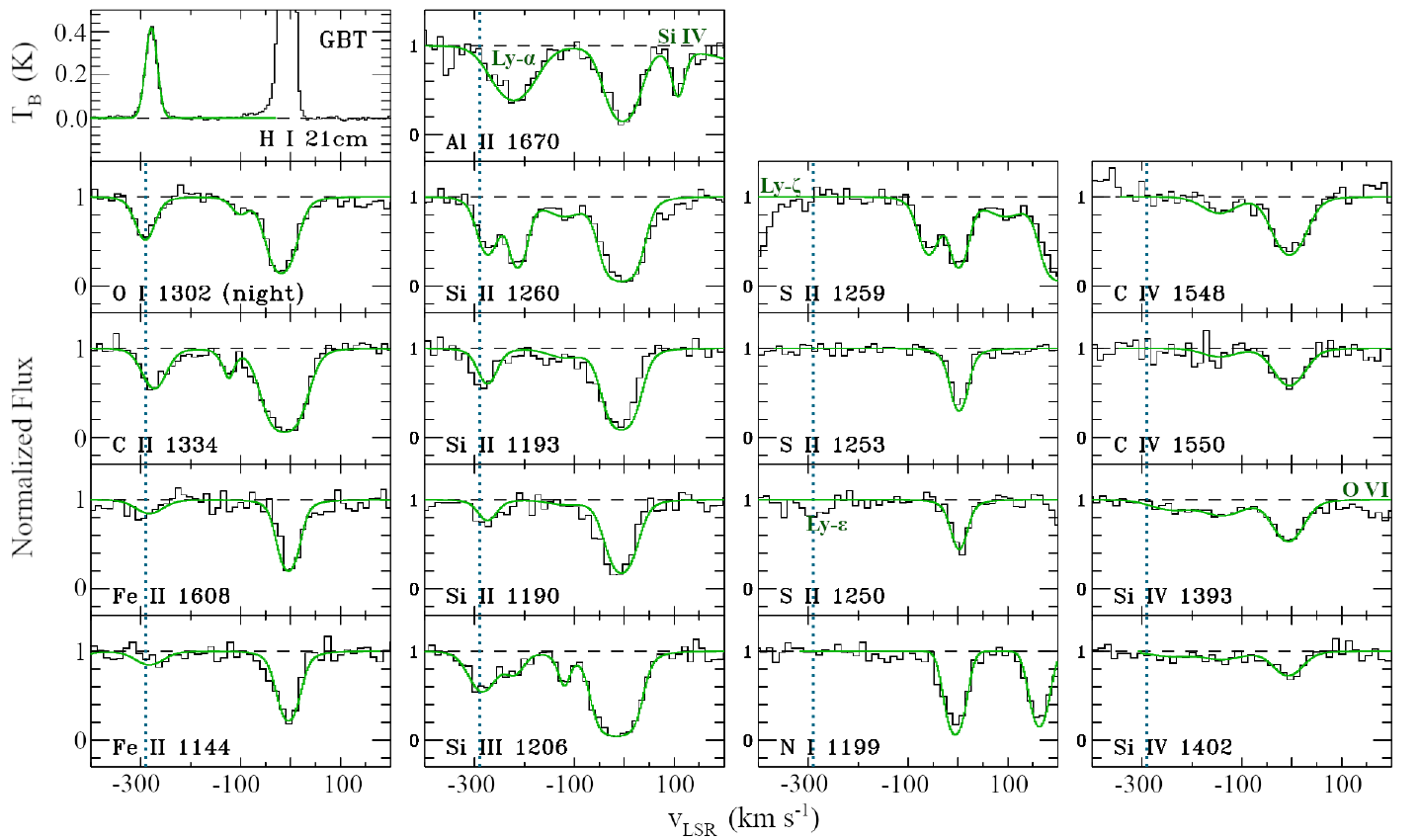}}\\
\caption{Continued from above.} 
\end{figure}

\section{Cloudy Results}\label{appendix:cloudy}

The full \textit{Cloudy} photoionization model results are shown in 
Table~\ref{table:cloudy-appen}, 
where the errors on each quantity are based on the \SiIII/\SiII\ measurement errors. Occasionally, \textit{Cloudy} was unable to produce a model for the density inferred by the $\pm1\sigma$ \SiIII/\SiII\ ratio (typically the +1$\sigma$ value for models with low $\Phi$ or higher distances). In these cases, the 1$\sigma$ log\,$n_{\rm H}$ was assumed to be symmetric about the measured  \SiIII/\SiII\ ratio. 

For the HVC toward HE\,0027, the modeled iron $\pm1\sigma$ IC errors were both larger than the measured IC for distances $\le75$ kpc. In these cases the log\,$n_{\rm H}$ grid resulted in a local minimum for iron IC near the matching \SiIII/\SiII\ measurements. 
For these cases, we chose the higher of the two errors on IC and assumed the error was symmetric about the IC from derived from the measured \SiIII/\SiII\ ratio. 

\begin{deluxetable}{lccccc}[!ht]
\tablecaption{Results from \textit{Cloudy} Photoionization Models to Small HVCs}\label{table:cloudy-appen}
\tablehead{\colhead{Sight Line}  & \colhead{Distance} & \colhead{IC(O)} & \colhead{[O/H]$_{\rm IC\,corr}$}   &  \colhead{Depth (kpc)} & \colhead{$P/k$ (cm$^{-3}$\,K)} }
\startdata
CTS\,47    & $5$ & $-0.0322\pm^{0.0056}_{0.0009}$ & $-1.36\pm0.13$ & $0.13\pm^{0.10}_{0.05}$ & $234\pm^{132}_{93}$\\
   & $10$ & $-0.0326\pm^{0.0057}_{0.0009}$ & $-1.36\pm0.13$ & $0.20\pm^{0.15}_{0.08}$ & $160\pm^{90}_{62}$\\
   & $20$ & $-0.0333\pm^{0.0057}_{0.0010}$ & $-1.36\pm0.13$ & $0.36\pm^{0.26}_{0.14}$ & $94\pm^{52}_{36}$\\
   & $50$ & $-0.0341\pm^{0.0055}_{0.0011}$ & $-1.36\pm0.13$ & $0.54\pm^{0.38}_{0.20}$ & $66\pm^{32}_{24}$\\
   & $75$ & $-0.0348\pm^{0.0061}_{0.0013}$ & $-1.36\pm0.13$ & $0.76\pm^{0.53}_{0.28}$ & $49\pm^{24}_{18}$\\
   & $100$ & $-0.0353\pm^{0.0060}_{0.0015}$ & $-1.36\pm0.13$ & $0.96\pm^{0.59}_{0.35}$ & $40\pm^{19}_{13}$\\
   & $150$ & $-0.0358\pm^{0.0061}_{0.0017}$ & $-1.36\pm0.13$ & $1.15\pm^{0.67}_{0.41}$ & $35\pm^{16}_{11}$\\
   & Extragalactic & $-0.0363\pm^{0.0063}_{0.0019}$ & $-1.36\pm0.13$ & $1.35\pm^{0.74}_{0.46}$ & $30\pm^{13}_{9}$\\   \hline
HE\,0027   & $5$ & $-0.0372\pm^{0.0705}_{0.0202}$ & $-0.80\pm^{0.19}_{0.17}$ & $0.03\pm^{0.04}_{0.01}$ & $553\pm^{527}_{303}$\\
   & $10$ & $-0.0373\pm^{0.0704}_{0.0201}$ & $-0.80\pm^{0.17}_{0.16}$ & $0.03\pm^{0.04}_{0.01}$ & $565\pm^{535}_{309}$\\
   & $20$ & $-0.0377\pm^{0.0698}_{0.0197}$ & $-0.80\pm^{0.17}_{0.16}$ & $0.04\pm^{0.06}_{0.02}$ & $377\pm^{353}_{205}$\\
   & $50$ & $-0.0373\pm^{0.0664}_{0.0192}$ & $-0.80\pm^{0.17}_{0.16}$ & $0.12\pm^{0.18}_{0.06}$ & $135\pm^{113}_{71}$\\
   & $75$ & $-0.0362\pm^{0.0706}_{0.0197}$ & $-0.80\pm^{0.17}_{0.16}$ & $0.19\pm^{0.33}_{0.10}$ & $84\pm^{71}_{47}$\\
   & $100$ & $-0.0349\pm0.0202$ & $-0.80\pm0.16$ & $0.26\pm0.14$ & $63\pm50$\\
   & $150$ & $-0.0323\pm0.0215$ & $-0.80\pm0.16$ & $0.38\pm0.19$ & $46\pm34$\\ 
   & Extragalactic & $-0.0283\pm0.0233$ & $-0.79\pm0.16$ & $0.59\pm0.28$ & $31\pm21$\\\hline
IRAS\,0459  &  $5$  & $-0.0075\pm^{0.0213}_{0.0158}$ & $-0.78\pm0.15$ & $0.13\pm^{0.04}_{0.03}$ & $129\pm^{26}_{23}$\\
   & $10$ & $-0.0026\pm^{0.0220}_{0.0181}$  & $-0.77\pm0.15$  & $0.21\pm^{0.06}_{0.05}$  & $81\pm^{18}_{15}$\\
   & $20$ & $-0.0053\pm^{0.0224}_{0.0161}$  & $-0.77\pm0.15$  & $0.36\pm^{0.12}_{0.07}$  & $49\pm10$\\
   & $50$ & $-0.0019\pm^{0.0205}_{0.0165}$  & $-0.77\pm0.15$  & $0.51\pm^{0.16}_{0.11}$  & $36\pm7$\\
   & $75$ & $0.0027\pm0.0171$  & $-0.77\pm0.15$  & $0.65\pm0.13$  & $29\pm6$\\
   & $100$ & $0.0015\pm0.0154$  & $-0.77\pm0.15$  & $0.71\pm0.13$  & $27\pm5$\\
   & $150$ & $0.0033\pm0.0172$  & $-0.77\pm0.15$  & $0.80\pm0.16$  & $24\pm5$\\   
   & Extragalactic & $0.0026\pm0.0141$  & $-0.77\pm0.15$  & $0.91\pm0.20$  & $21\pm6$\\ \hline
Mrk\,969 &  $5$  & $-0.0270\pm^{0.0055}_{0.0045}$ & $-1.69\pm0.15$  & $0.11\pm0.02$  & $231\pm^{45}_{37}$\\ 
   & $10$ & $-0.0273\pm^{0.0061}_{0.0040}$ & $-1.69\pm0.15$ & $0.12\pm^{0.03}_{0.02}$ & $212\pm^{36}_{38}$\\
   & $20$ & $-0.0274\pm^{0.0054}_{0.0040}$ & $-1.69\pm0.15$ & $0.17\pm^{0.04}_{0.03}$ & $150\pm^{26}_{24}$\\
   & $50$ & $-0.0276\pm^{0.0056}_{0.0045}$ & $-1.69\pm0.15$ & $0.44\pm^{0.10}_{0.08}$ & $64\pm^{12}_{10}$\\
   & $75$ & $-0.0274\pm^{0.0050}_{0.0047}$ & $-1.69\pm0.15$ & $0.68\pm^{0.14}_{0.00}$ & $44\pm^{8}_{6}$\\
   & $100$ & $-0.0275\pm^{0.0050}_{0.0043}$ & $-1.69\pm0.15$ & $0.87\pm^{0.18}_{0.15}$ & $35\pm^{6}_{5}$\\
   & $150$ & $-0.0281\pm^{0.0050}_{0.0039}$ & $-1.69\pm0.15$ & $1.12\pm^{0.23}_{0.17}$ & $28\pm4$\\ 
  & Extragalactic & $-0.0280\pm0.0038$ & $-1.69\pm0.15$ & $1.59\pm0.20$ & $22\pm3$\\ \hline
UVQS\,J0110 &  $5$   & $-0.0164\pm0.0004$ & $-1.72\pm0.13$ & $0.33\pm^{0.14}_{0.09}$ & $271\pm^{83}_{72}$\\
   & $10$ & $-0.0166\pm0.0004$ & $-1.72\pm0.13$ & $0.36\pm^{0.15}_{0.09}$ & $249\pm^{77}_{66}$\\
   & $20$ & $-0.0176\pm0.0002$ & $-1.72\pm0.13$ & $0.53\pm^{0.22}_{0.14}$ & $179\pm^{55}_{47}$\\
   & $50$ & $-0.0212\pm^{0.0004}_{0.0000}$ & $-1.72\pm0.13$ & $1.38\pm^{0.53}_{0.36}$ & $77\pm^{23}_{19}$\\
   & $75$ & $-0.0234\pm^{0.0004}_{0.0002}$ & $-1.72\pm0.13$ & $2.19\pm^{0.76}_{0.57}$ & $52\pm^{15}_{12}$\\
   & $100$ & $-0.0250\pm^{0.0005}_{0.0002}$ & $-1.72\pm0.13$ & $2.89\pm^{1.10}_{0.70}$ & $41\pm^{11}_{10}$\\
   & $150$ & $-0.0270\pm^{0.0005}_{0.0002}$ & $-1.73\pm0.13$ & $3.99\pm^{1.27}_{0.89}$ & $31\pm7$\\ 
   & Extragalactic & $-0.0298\pm0.0004$ & $-1.73\pm0.13$ & $5.90\pm{1.10}$ & $23\pm4$\\ 
\enddata
\tablecomments{For each HVC and for each distance modeled, this table reports the oxygen ionization correction, the corrected oxygen abundance, the line-of-sight cloud depth, and the thermal pressure.}
\end{deluxetable}

\section{Beam Smearing}\label{appendix:beam_smearing}

The 9\arcmin\ GBT 21 cm beam probes a much larger region of gas than the HST/COS pencil beam. As such, the median \HI\ column density measured over the GBT beam is likely not representative of the true \HI\ column density along the COS sightline. Here we use the high- and low-resolution \HI\ data (9\arcmin\ and 2\arcmin) of three HVCs presented in \citet{Faridani_2014} to estimate the error in our \HI\ column measurements. While this technique does not give us an exact measurement of the beam smearing between a 9\arcmin\ and pencil-beam, it provides a quantitative measurement of the expected beam-smearing effects as the resolution of the \HI\ data decreases.

\citet{Faridani_2014} present \HI\ data for 3 Compact HVCs (CHVC\,070+51-150, CHVC\,108-21-390, and CHVC\,162+03-186) from the Effelsberg telescope and the Westerbork Synthesis Radio Telescope (WSRT). The Effelsberg 21cm beam is approximately the same size as the GBT \HI\ beam at 9\arcmin, while the WSRT provides \HI\ data with a beamsize of \s2\arcmin. For each CHVC \citet{Faridani_2014} also feather the low- and high-resolution data together to capture the low-density components measured with the Effelsberg telescope and the high-resolution, high-density emission measured with the WSRT \citep[using the methods described in][]{Faridani_2018}. The resulting feathered data have the same synthesized beamsize as the WSRT data. 

\begin{figure}[!ht]
   \centering
    \begin{tabular}{cc} 
    \raisebox{0.1\height}{\includegraphics[width=0.3\textwidth]{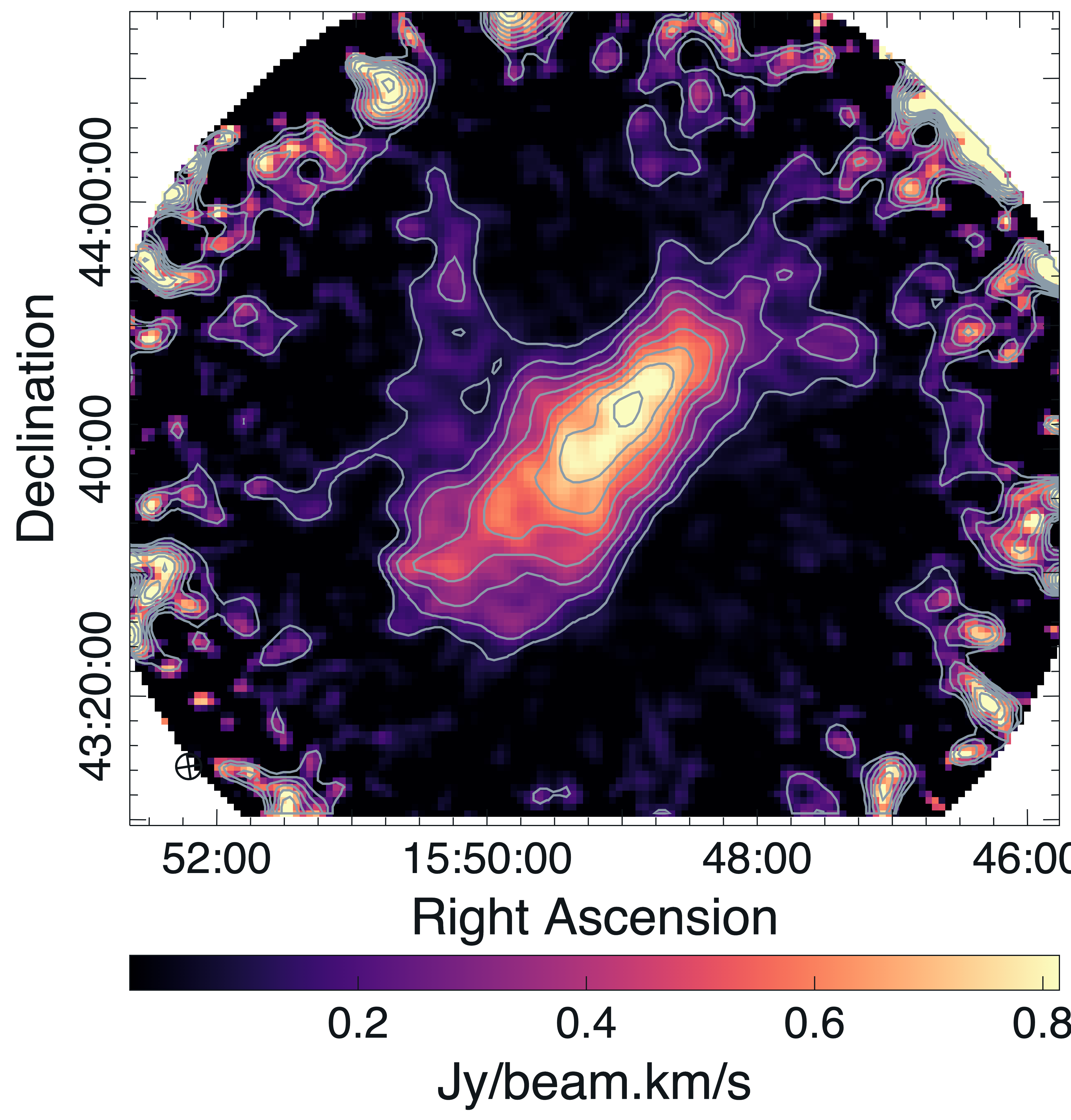}}  &      \includegraphics[width=0.4\textwidth]{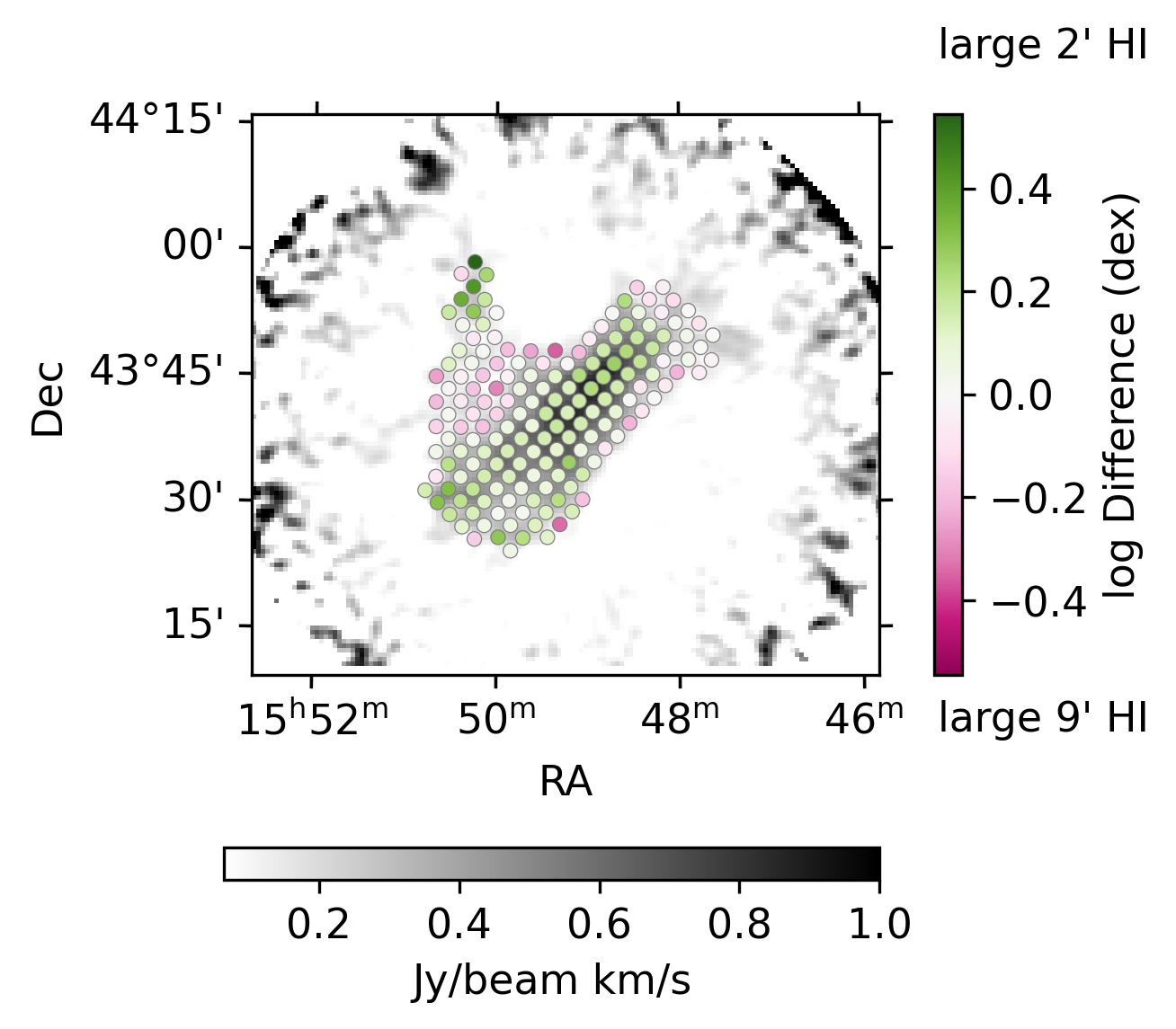} \\   \raisebox{0.1\height}{\includegraphics[width=0.3\textwidth]{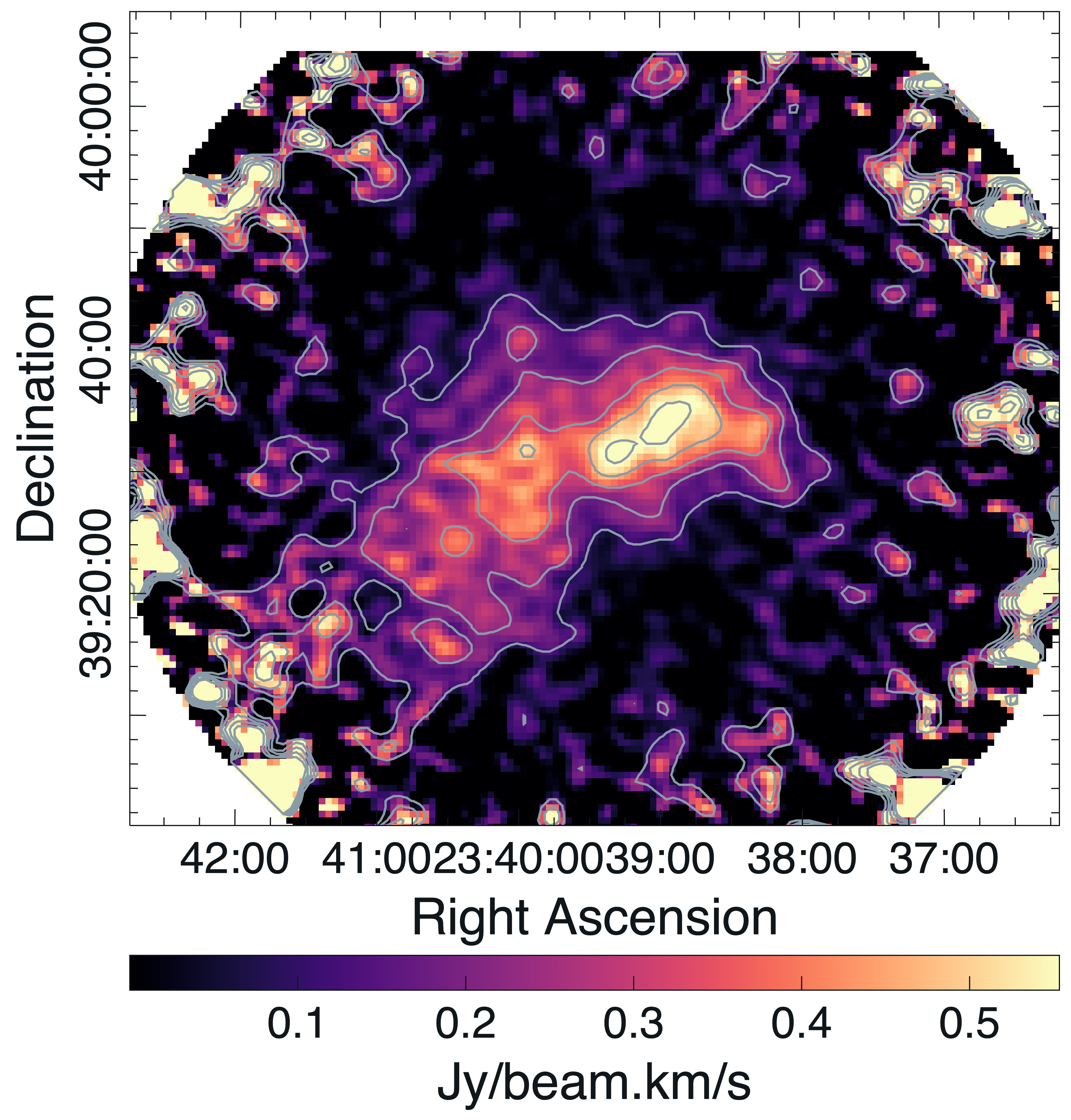}} & \includegraphics[width=0.4\textwidth]{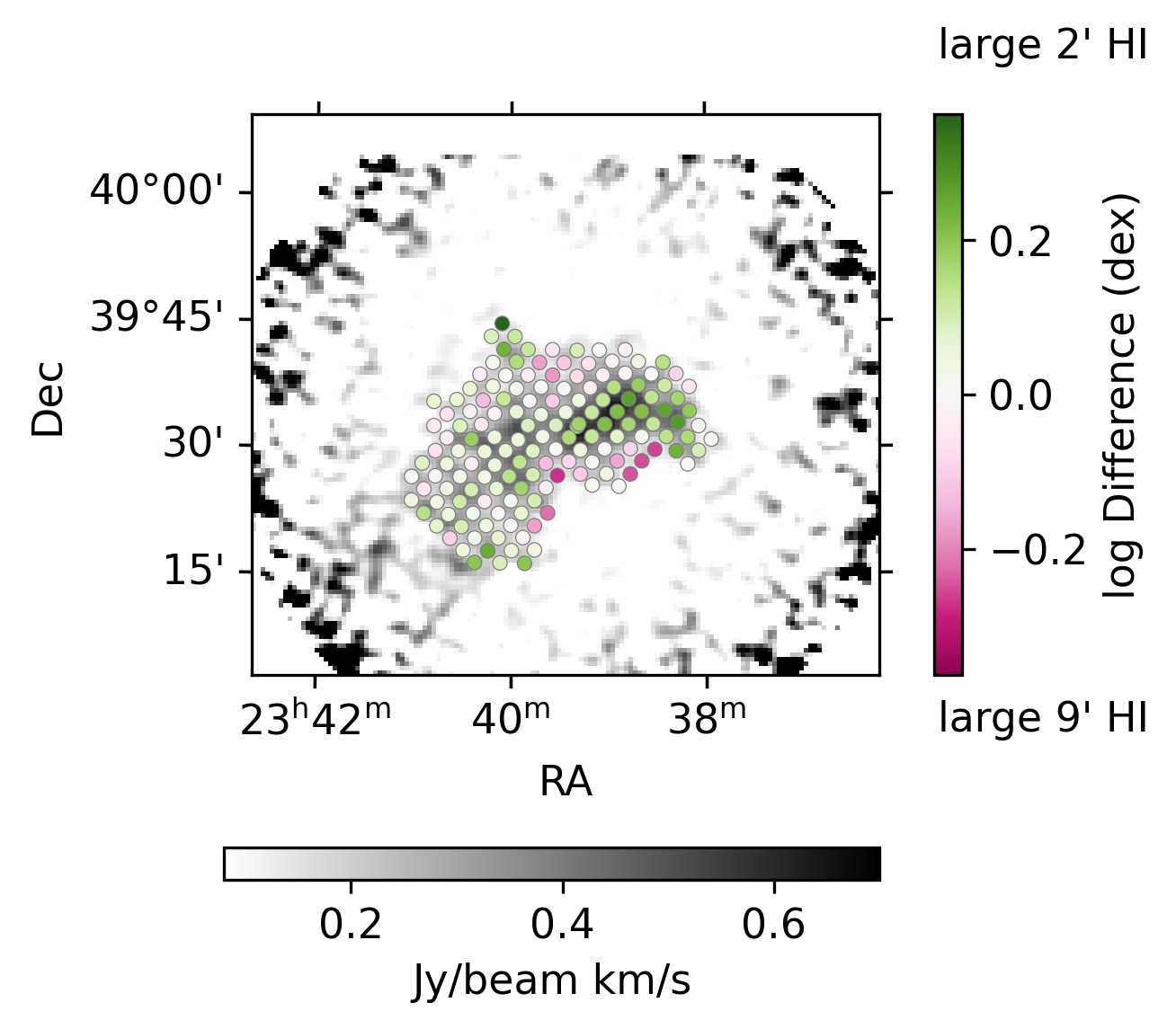}   \\ \raisebox{0.1\height}{\includegraphics[width=0.3\textwidth]{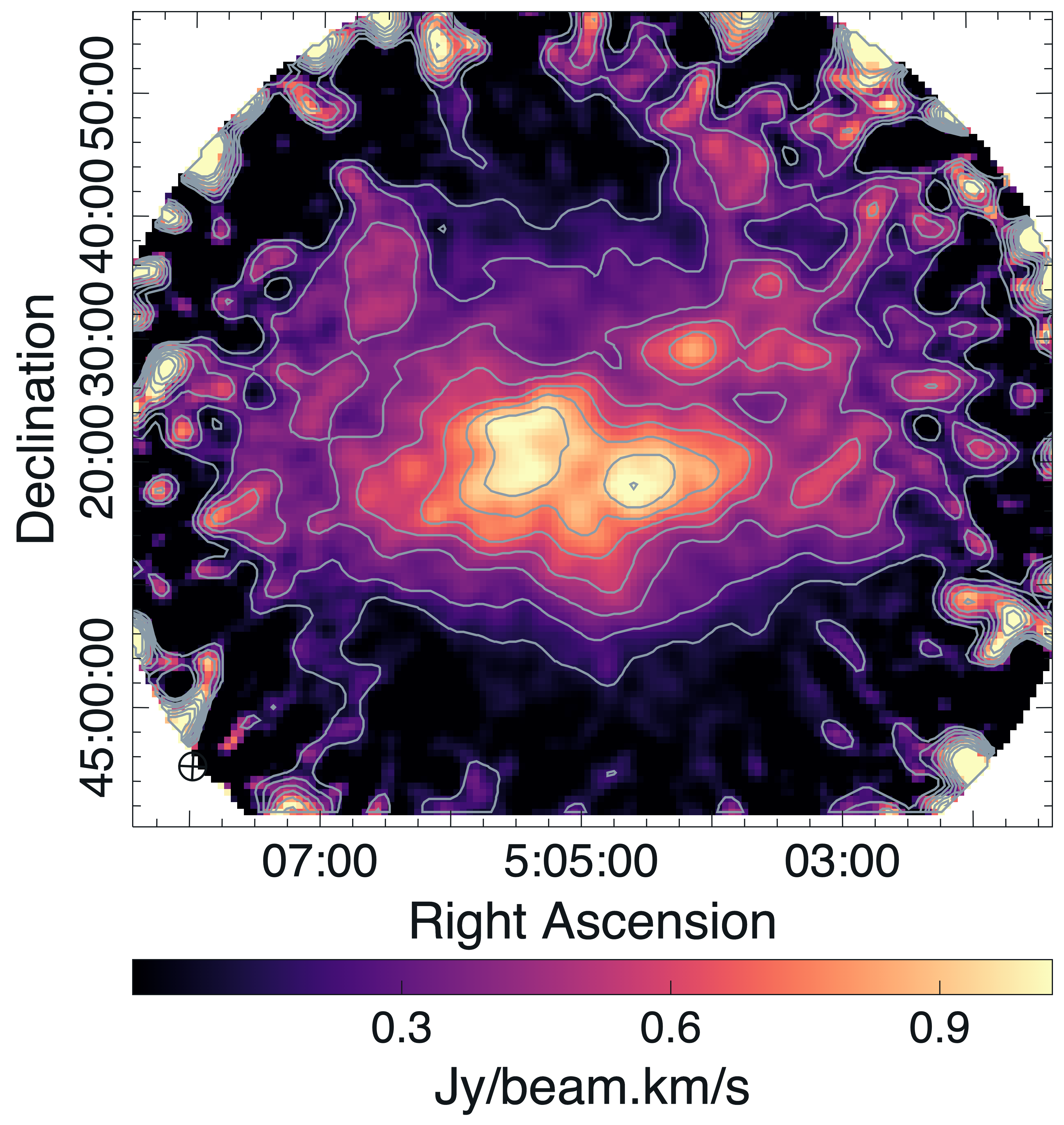}} &
    \includegraphics[width=0.4\textwidth]{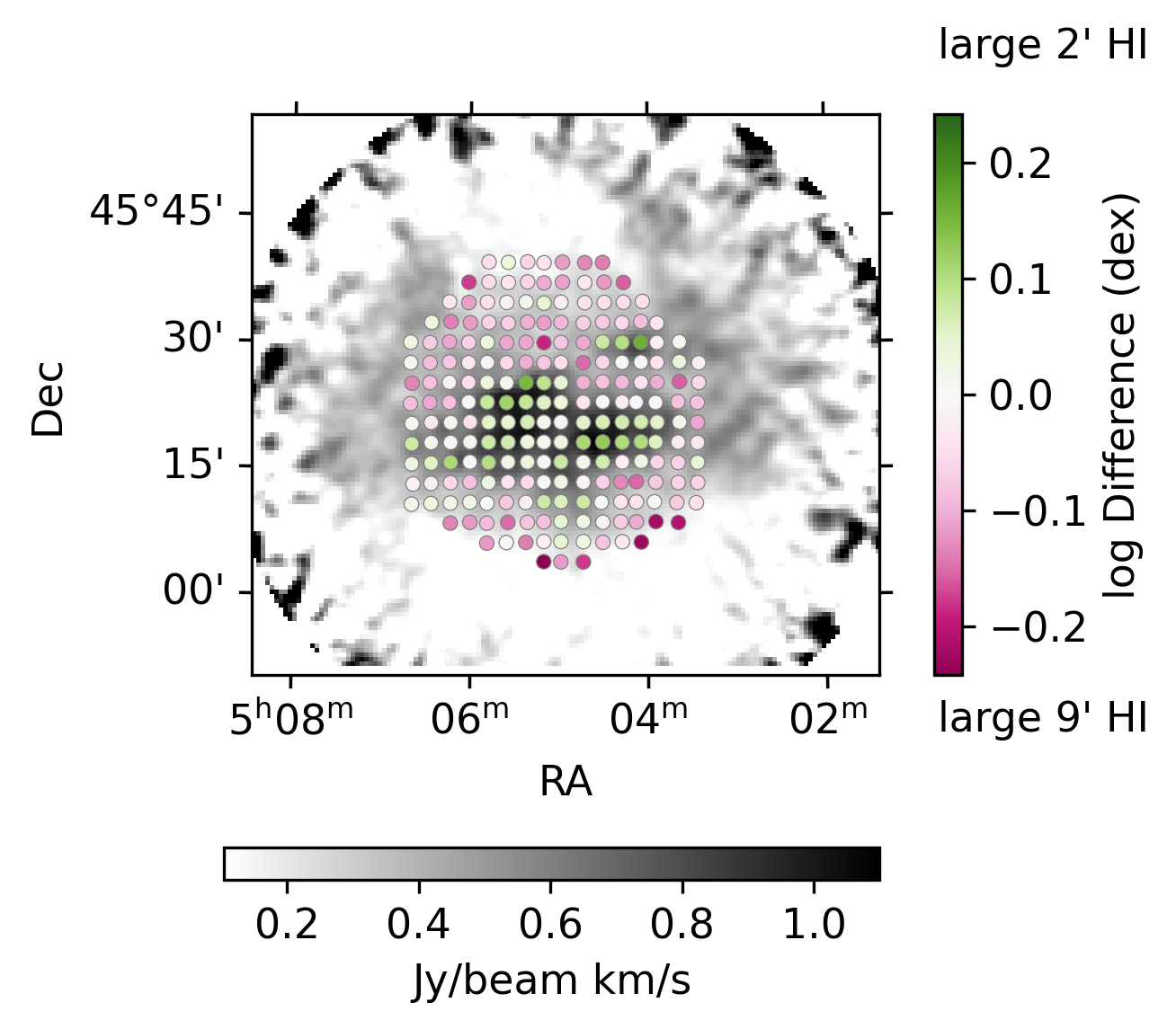}\\
\end{tabular}
\caption{Left panels: zeroth moment \HI\ maps of the feathered WSRT and Effelsberg data for the three CHVCs in \citet{Faridani_2014}. Contours are as follows: CHVC 070+51: 0.051$\times$(2, 6, 10, 14, 18, 22, 26, 30) Jy\,beam$^{-1}$\,\kms; CHVC 108-21: 0.037$\times$(3, 6, 9, 12, 15) Jy\,beam$^{-1}$\,\kms; CHVC 162+03: 0.032$\times$(4, 8, 12, 16, 20, 24, 28, 32) Jy\,beam$^{-1}$\,\kms. Right panels: Greyscale zeroth moment \HI\ maps and circles indicating the locations of \HI\ measurements in the two different map resolutions. The colors of the circles represent the logarithmic difference in the \HI\ columns measured at the two resolutions, in the sense of the 9\arcmin\ $N$(\HI) subtracted from 2\arcmin\ $N$(\HI).}
\label{figure:beam_smearing-appendix}
\end{figure}

For each of the CHVCs in the \citet{Faridani_2014} sample, we measured the \HI\ column density in both the feathered data cube (2\arcmin) and the Effelsberg data cube (9\arcmin). We use the feathered data since it is the most accurate high-resolution representation of the total \HI\ associated with the CHVCs and we compare to the Effelsberg data since those have very similar resolution to our own GBT observations. We used CARTA \citep{Wang_2020} to make zeroth moment maps of each feathered data cube and placed a grid of ellipse regions (sized to each individual data cube's synthesized beam) over the emission associated with the CHVC (out to \s2-4$\sigma$). The zeroth moment maps and the respective locations of the ellipses can be seen in Figure~\ref{figure:beam_smearing-appendix}. We then created a grid of circles with the same central locations as those in the feathered data cubes, but with a diameter of 9\arcmin\ for the Effelsberg data cube. Each of the zeroth-moment feathered maps have large noise fluctuations around the edges; therefore, we removed any regions where the 9\arcmin\ circle encompassed a significant portion of this noise. 

For each ellipse or circle, we obtained an \HI\ spectrum from the respective data cubes (\s2\arcmin\ ellipses for the feathered data cube and 9\arcmin\ circles for the Effelsberg data cube) 
and fit a Gaussian(s) to the spectrum. If a spectrum had more than one Gaussian component in either data cube, then those components were checked against the components measured in the other data cube from the region of the same central location. Only components that covered similar velocity ranges in both data cubes and could reasonably be assumed to be part of the same cloud were kept in the analysis. For each ellipse or circle, we added all Gaussian components. Then, for each circle or ellipse with the same central location, we compared the \s2\arcmin\ feathered to the 9\arcmin\ Effelsberg total \HI\ column measurements. The \HI\ column densities were calculated using Equation~\ref{equation:nhi} and the equation for relating brightness temperature to flux (S): 
\begin{equation}
T_{B} = \frac{S\,[\mathrm{mJy\:beam}^{-1}]}{1.65\times10^{-3} \Delta\alpha\Delta\delta},
\end{equation}
\noindent where $\Delta\alpha$ and $\Delta\delta$ are the FWHM values along the beam's major and minor axes in arcsec. 
In Figure~\ref{figure:beam_smearing-appendix} we plotted the logarithmic difference in the \HI\ columns measured at the two resolutions: 
log\,$N$(\HI)$_{2\arcmin}$ - log\,$N$(\HI)$_{9\arcmin}$.
Positive differences (green) represent a higher \HI\ measurement in the 2\arcmin\ beam (feathered data cube) and negative difference (pink) represent a higher \HI\ measurement in the 9\arcmin\ beam (Effelsberg data cube). 

\begin{figure}[!ht]
   \centering
   \epsscale{0.5}
   \plotone{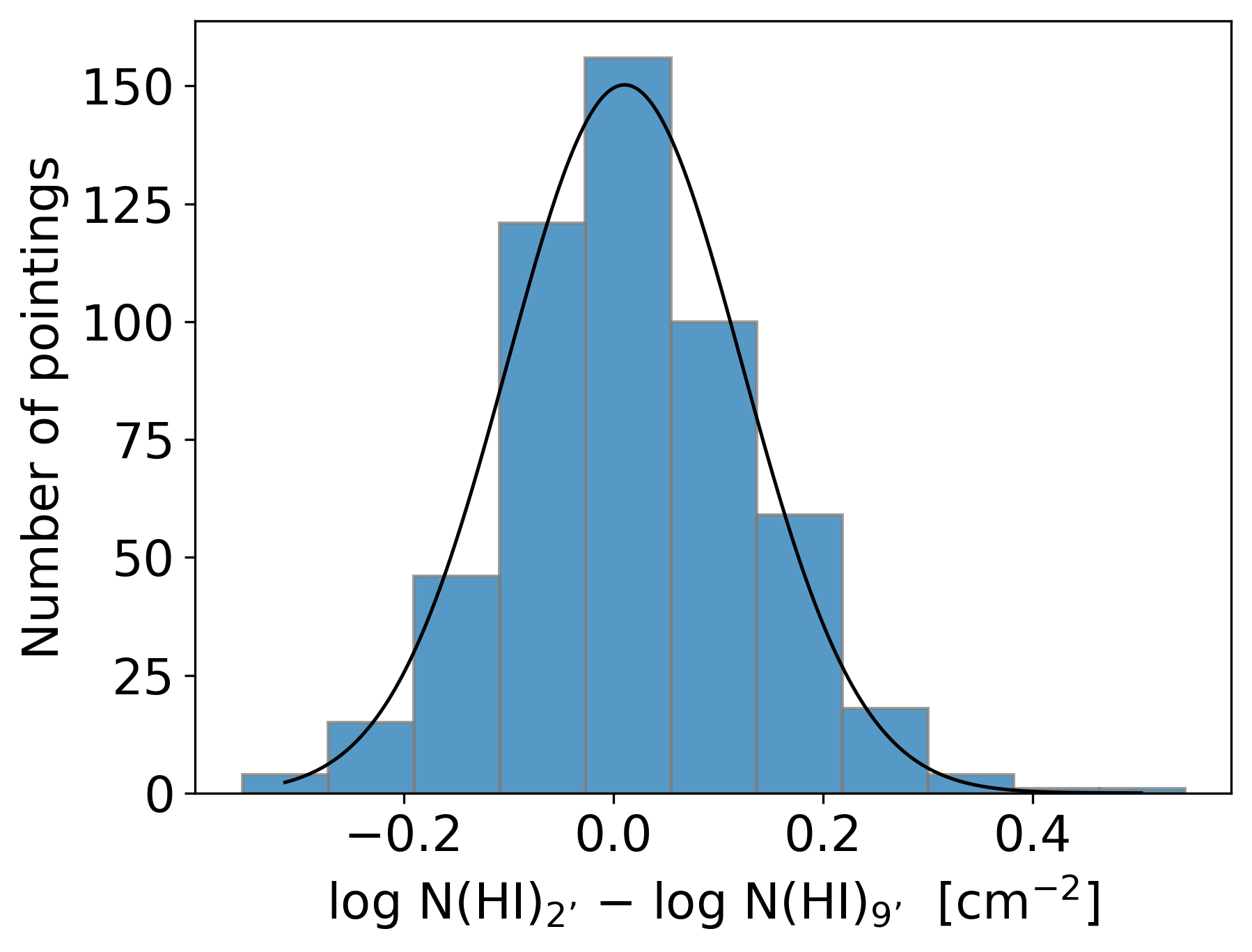}
\caption{The distribution of the difference in \HI\ column density between the feathered and low-resolution \HI\ maps. Negative numbers represent a larger \HI\ column in the 9\arcmin\ data and positive numbers represent a larger \HI\ column in the 2\arcmin\ data.}
\label{figure:log_diff_histogram-appendix}
\end{figure}

A histogram showing the logarithmic difference for all of the compared regions in all three CHVCs is shown in Figure~\ref{figure:log_diff_histogram-appendix}. We fit a Gaussian function to this distribution and found a 1$\sigma$ dispersion of 0.11 dex. This number is used throughout the paper as the systematic beam-smearing error on our \HI\ measurements. This value is close to 0.15 dex, which is often considered a reasonable estimate on HVC beam-smearing errors \citep{Fox_2018, Ashley_2022}. 

In the case of our low-metallicity measurements, the highest difference between the 2\arcmin\ and 9\arcmin\ data cubes of 0.54 dex increases the error on the linear metallicity measurement to: $0.044\pm^{0.055}_{0.044}$ $Z_\sun$, $0.021\pm^{0.026}_{0.021}$ $Z_\sun$, and $0.019\pm^{0.024}_{0.019}$ $Z_\sun$ for CTS\,47, Mrk\,969, and UVQS\,J0110, respectively (see Table~\ref{table:metallicities}). Therefore, even in the event that strong beam-smearing causes all our sample HVCs to have an overestimated \HI\ column density (and therefore an underestimated metallicity), the metallicities of the CTS\,47, Mrk\,969, and UVQS\,J0110 HVCs are still $\lesssim$0.10 $Z_\sun$ and both Mrk\,969 and UVQS\,J0110 retain metallicities \textless0.05 $Z_\sun$. These low metallicities are therefore robust against beam-smearing effects.

\begin{figure}[!ht]
   \centering
   \epsscale{0.6}
   \plotone{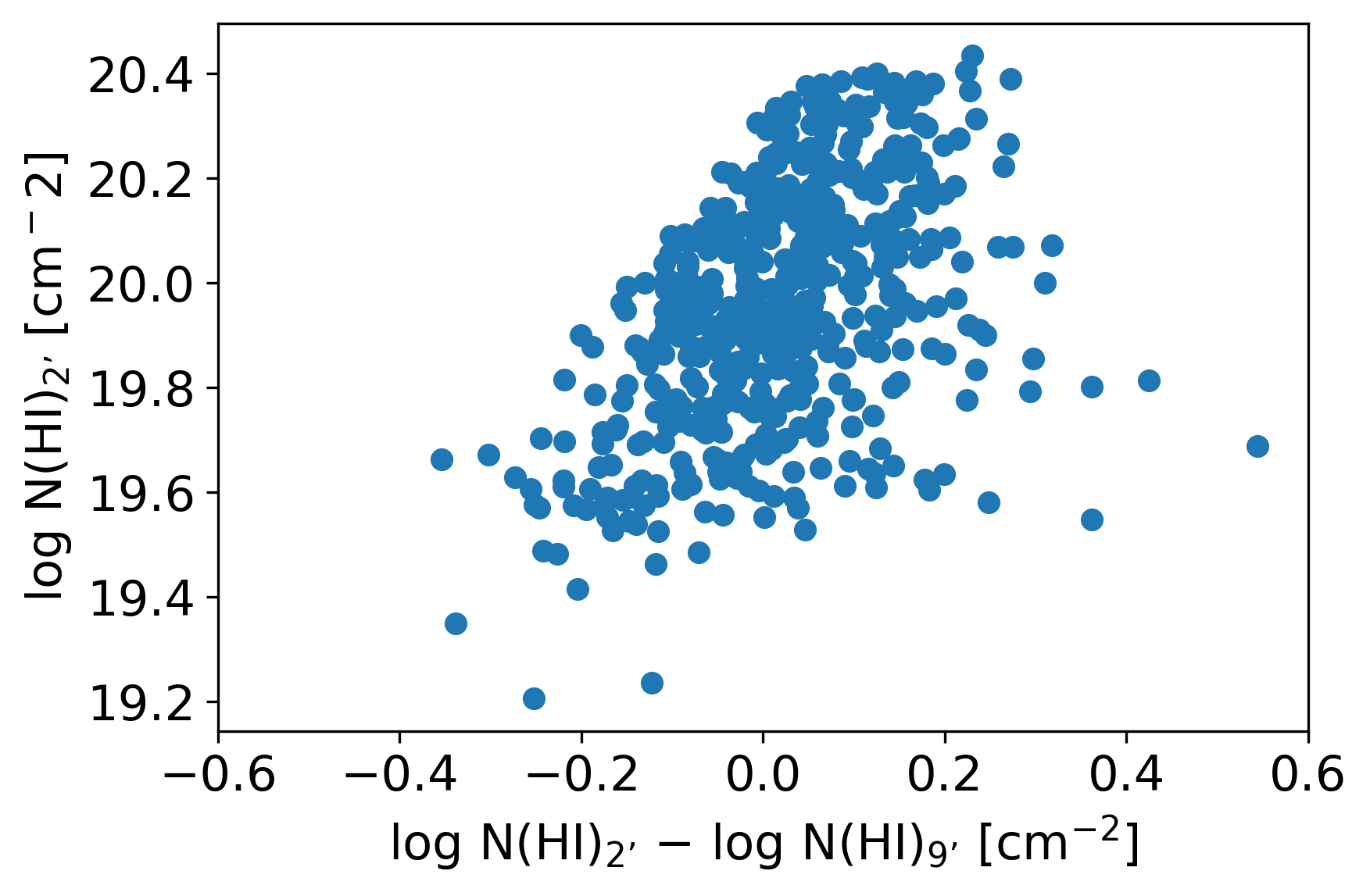}
\caption{The relationship between the \HI\ column density measured in the 2\arcmin\ feathered datacube and the difference between the \HI\ column densities measured in the 9\arcmin\ data cube and the 2\arcmin\ feathered data cube. Negative numbers on the x-axis represent a larger \HI\ column in the 9\arcmin\ data and positive numbers represent a larger \HI\ column in the 2\arcmin\ data.}
\label{figure:HI_column_vs_diff-appendix}
\end{figure}

We also plot the logarithmic difference between the 2\arcmin\ and 9\arcmin\ \HI\ maps against 
the 2\arcmin\ \HI\ columns (which provides the most reliable \HI\ measurements)
in Figure~\ref{figure:HI_column_vs_diff-appendix}. We see that at low $N$(\HI), the 9\arcmin\ data  tends to overestimate the \HI\ column, whereas at high $N$(\HI), the 9\arcmin\ data tends to underestimate the \HI\ column density. As such, when using the larger 9\arcmin\ beam of the GBT, sight lines that pass through the low $N$(\HI) regions of the CHVC are more likely to \emph{underestimate} the metallicity 
and sight lines that pass through in the high $N$(\HI) regions of the CHVC are more likely to \emph{overestimate} the metallicity. 
While there is significant scatter in this relationship, each of the three background AGNs that result in a low-metallicity measurement (CTS\,47, Mrk\,969, and UVQS\,J0110) pass through a central, higher-density region in their respective HVC, providing us with confidence that these three HVCs have genuinely low metallicities.

\section{Dust Depletion}\label{appendix:dust}

We list the full dust depletions for each measured ion and assumed distance in Table~\ref{table:depletions-appendix}.

\begin{deluxetable}{lccccccc}
\tablecaption{HVC Dust Depletion Measurements}
\setlength{\tabcolsep}{4pt}
\tablehead{\colhead{Sight Line}  & \colhead{$d$ (kpc)} & \colhead{$\delta_{\rm O}$(S)} & \colhead{$\delta_{\rm O}$(N)}  & \colhead{$\delta_{\rm O}$(C)}   & \colhead{$\delta_{\rm O}$(Si)} & \colhead{$\delta_{\rm O}$(Al)} & \colhead{$\delta_{\rm O}$(Fe)}  }   
\startdata
CTS\,47 & $5$ & $<1.22$ & $<0.65$ & $-0.65\pm^{0.16}_{0.15}$ & $-0.45\pm0.12$  & $-0.68\pm^{0.29}_{0.25}$ & $<0.62$ \\ 
 & $10$ & $<1.23$ & $<0.65$ & $-0.65\pm^{0.17}_{0.15}$ & $-0.44\pm0.12$  & $-0.67\pm^{0.29}_{0.25}$ & $<0.63$ \\ 
 & $20$ & $<1.23$ & $<0.66$ & $-0.64\pm^{0.18}_{0.16}$ & $-0.42\pm0.12$  & $-0.64\pm^{0.29}_{0.25}$ & $<0.64$ \\ 
 & $50$ & $<1.25$ & $<0.67$ & $-0.62\pm^{0.19}_{0.15}$ & $-0.40\pm^{0.12}_{0.11}$  & $-0.59\pm^{0.28}_{0.23}$ & $<0.66$  \\
 & $75$ & $<1.26$ & $<0.68$ & $-0.61\pm^{0.21}_{0.15}$ & $-0.39\pm^{0.12}_{0.11}$  & $-0.54\pm^{0.29}_{0.23}$ & $<0.66$ \\ 
 & $100$ & $<1.27$ & $<0.70$ & $-0.59\pm^{0.21}_{0.16}$ & $-0.38\pm^{0.12}_{0.11}$  & $-0.49\pm^{0.28}_{0.23}$ & $<0.66$ \\ 
 & $150$ & $<1.29$ & $<0.71$ & $-0.57\pm^{0.22}_{0.15}$ & $-0.38\pm^{0.12}_{0.11}$  & $-0.44\pm^{0.28}_{0.22}$ & $<0.66$  \\
 & Extragalactic & $<1.30$ & $<0.72$ & $-0.55\pm^{0.15}_{0.24}$ & $-0.37\pm^{0.11}_{0.12}$  & $-0.39\pm^{0.21}_{0.29}$ & $<0.66$ \\ \hline 
HE\,0027 & $5$ & $<0.32$ & $<0.12$ & $-0.70\pm^{0.25}_{0.24}$ & $-0.30\pm^{0.20}_{0.18}$  & $-0.63\pm^{0.42}_{0.49}$ & $0.08\pm^{0.16}_{0.15}$  \\
 & $10$ & $<0.32$ & $<0.12$ & $-0.70\pm^{0.25}_{0.24}$ & $-0.39\pm^{0.21}_{0.19}$  & $-0.32\pm^{0.42}_{0.49}$ & $0.08\pm^{0.16}_{0.15}$ \\ 
 & $20$ & $<0.32$ & $<0.12$ & $-0.70\pm0.25$ & $-0.37\pm^{0.20}_{0.18}$  & $-0.62\pm^{0.43}_{0.51}$ & $0.09\pm^{0.17}_{0.15}$ \\ 
 & $50$ & $<0.33$ & $<0.12$ & $-0.69\pm^{0.25}_{0.31}$ & $-0.31\pm^{0.18}_{0.19}$  & $-0.60\pm^{0.42}_{0.54}$ & $0.15\pm^{0.17}_{0.16}$   \\
 & $75$ & $<0.33$ & $<0.12$ & $-0.68\pm^{0.26}_{0.44}$ & $-0.28\pm^{0.18}_{22}$  & $-0.56\pm^{0.42}_{0.61}$ & $0.18\pm^{0.17}_{0.16}$ \\ 
 & $100$ & $<0.34$ & $<0.12$ & $-0.67\pm0.26$ & $-0.26\pm0.17$  & $-0.51\pm0.41$ & $0.20\pm0.15$ \\ 
 & $150$ & $<0.36$ & $<0.12$ & $-0.64\pm0.26$ & $-0.23\pm0.16$  & $-0.44\pm0.39$ & $0.22\pm0.15$ \\ 
 & Extragalactic & $<0.4$ & $<0.13$ & $-0.59\pm0.26$ & $-0.20\pm0.16$  & $-0.30\pm0.37$ & $0.25\pm0.15$ \\ \hline
IRAS\,0459 & $5$ & $<0.21$ & $<0.30$ 
& $-0.99\pm0.14$ & $-0.82\pm0.12$  & $-0.92\pm^{0.18}_{0.19}$ & $<0.08$ \\ 
 & $10$ & $<0.19$ & $<0.30$ 
 & $-1.02\pm0.15$ & $-0.81\pm^{0.12}_{0.13}$  & $-0.94\pm^{0.19}_{0.20}$ & $<0.11$ \\ 
 & $20$ & $<0.20$ & $<0.28$ 
 & $-1.01\pm^{0.15}_{0.19}$ & $-0.76\pm^{0.12}_{0.13}$  & $-0.85\pm^{0.19}_{0.22}$ & $<0.16$ \\ 
 & $50$ & $<0.19$ & $<0.26$ 
 & $-1.03\pm^{0.17}_{0.25}$ & $-0.74\pm^{0.13}_{0.14}$  & $-0.81\pm^{0.20}_{0.26}$ & $<0.18$ \\ 
 & $75$ & $<0.16$ & $<0.24$ 
 & $-1.07\pm0.19$ & $-0.74\pm0.13$  & $-0.80\pm0.21$ & $<0.19$ \\ 
 & $100$ & $<0.17$ & $<0.24$ 
 & $-1.05\pm0.19$ & $-0.72\pm0.13$  & $-0.74\pm0.21$ & $<0.20$ \\ 
 & $150$ & $<0.15$ & $<0.22$ 
 & $-1.07\pm0.22$ & $-0.72\pm0.13$  & $-0.73\pm0.23$ & $<0.21$ \\ 
 & Extragalactic & $<0.16$ & $<0.22$ 
 & $-1.05\pm0.21$ & $-0.70\pm0.13$  & $-0.68\pm0.22$ & $<0.21$ \\ \hline
Mrk\,969 & $5$ & $<1.29$ & $<1.10$ & $-0.57\pm0.13$ & $-0.28\pm0.11$  & $-0.31\pm0.17$ & $<1.02$ \\ 
 & $10$ & $<1.29$ & $<1.10$ & $-0.56\pm0.13$ & $-0.27\pm0.11$  & $-0.31^{0.16}_{0.17}$ & $<1.03$ \\ 
 & $20$ & $<1.29$ & $<1.10$ & $-0.57\pm0.13$ & $-0.26\pm0.11$  & $-0.3\pm^{0.16}_{0.17}$ & $<1.04$ \\ 
 & $50$ & $<1.28$ & $<1.11$ & $-0.58\pm^{0.13}_{0.14}$ & $-0.23\pm0.11$  & $-0.27\pm^{0.17}_{0.18}$ & $<1.06$ \\ 
& $75$ & $<1.28$ & $<1.11$ & $-0.58\pm0.14$ & $-0.21\pm0.11$ & $-0.23\pm0.17$ & $<1.08$ \\
 & $100$ & $<1.28$ & $<1.11$ & $-0.58\pm^{0.14}_{0.16}$ & $-0.20\pm^{0.11}_{0.12}$  & $-0.18^{0.17}_{0.18}$ & $<1.09$ \\ 
 & $150$ & $<1.29$ & $<1.12$ & $-0.56\pm^{0.14}_{0.18}$ & $-0.18\pm^{0.11}_{0.12}$  & $-0.10\pm^{0.17}_{0.20}$ & $<1.09$ \\ 
 & Extragalactic & $<1.30$ & $<1.13$ & $-0.55\pm0.15$ & $-0.16\pm0.11$  & $0.01\pm^{0.14}_{0.17}$ & $<1.10$ \\ \hline
UVQS\,J0110 & $5$ & $<0.97$ & $<-0.21$ & $-0.61\pm^{0.10}_{0.11}$ & $-0.32\pm0.09$  & $<-0.24$ & $<0.63$ \\ 
 & $10$ & $<0.97$ & $<-0.21$ & $-0.61\pm^{0.10}_{0.11}$ & $-0.32\pm0.09$  & $<-0.25$ & $<0.63$ \\ 
 & $20$ & $<0.97$ & $<-0.21$ & $-0.61\pm^{0.10}_{0.11}$ & $-0.32\pm0.09$  & $<-0.24$ & $<0.63$ \\ 
 & $50$ & $<0.97$ & $<-0.19$ & $-0.62\pm^{0.11}_{0.12}$ & $-0.31\pm0.09$  & $<-0.21$ & $<0.63$ \\ 
 & $75$ & $<0.97$ & $<-0.18$ & $-0.62\pm^{0.11}_{0.12}$ & $-0.30\pm0.09$  & $<-0.17$ & $<0.62$ \\ 
 & $100$ & $<0.98$ & $<0.17$ & $-0.62\pm^{0.11}_{0.14}$ & $-0.31\pm0.09$  & $<-0.13$ & $<0.62$ \\ 
 & $150$ & $<0.99$ & $<-0.16$ & $-0.61\pm^{0.11}_{0.15}$ & $-0.31\pm0.09$  & $<-0.08$ & $<0.61$ \\ 
 & Extragalactic & $<1.00$ & $<-0.14$ & $-0.60\pm0.11$ & $-0.31\pm0.09$  & $<0.02$ & $<0.60$ \\ 
\enddata
\tablecomments{For each HVC and for each distance modeled, this table reports the depletion of sulfur, carbon, silicon, iron, aluminum, and nitrogen with respect to oxygen.}
\label{table:depletions-appendix}
\end{deluxetable}

\end{document}